\documentclass[12pt,preprint]{aastex}
\usepackage{graphicx,psfrag}
\newcommand\mathnew{\mathsurround=0pt}
\def\simov#1#2{\lower .5pt\vbox{\baselineskip0pt \lineskip-.5pt
        \ialign{$\mathnew#1\hfil##\hfil$\crcr#2\crcr\sim\crcr}}}

\newcommand{\msun}{${\rm M}_\odot$}

\def\lesssim{\mathrel{\hbox{\rlap{\hbox{\lower4pt\hbox{$\sim$}}}\hbox{$<$}}}
}
\def\la{\mathrel{\hbox{\rlap{\hbox{\lower4pt\hbox{$\sim$}}}\hbox{$<$}}}}
\def\ga{\mathrel{\hbox{\rlap{\hbox{\lower4pt\hbox{$\sim$}}}\hbox{$>$}}}}

\newcommand{\rxte}{{\it Rossi X-ray Timing Explorer\, }}

\newcommand{\RXTE}{{\it RXTE\,}}

\shortauthors{Wen  et al.}
\shorttitle{Periodicity Search in RXTE/ASM data}

\begin{document}

\title{A Systematic Search for Periodicities in RXTE/ASM Data}
\author{Linqing Wen\altaffilmark{1}, Alan M. Levine\altaffilmark{2},
  Robin H. D.   Corbet\altaffilmark{3}, and Hale V. Bradt\altaffilmark{2}}
\altaffiltext{1}{Max Planck Institut f\"{u}r Gravitationsphysik,
  Albert-Einstein-Institut, Am M\"{u}hlenberg 1, D-14476 Golm, Germany; lwen@aei.mpg.de}
\altaffiltext{2} {Center for Space Research, MIT, Cambridge, MA 02139,
  USA; aml@space.mit.edu, hale@space.mit.edu}
\altaffiltext{3}{Universities Space Research Association and X-ray Astrophysics Laboratory, Code 662,
NASA/Goddard Space Flight Center, Greenbelt, MD 20771, USA; corbet@gsfc.nasa.gov}

\begin{abstract}

We present the results of a systematic search in 8.5 years of {\it
Rossi X-ray Timing Explorer} All-Sky Monitor data for evidence of
periodicities.  The search was conducted by application of the
Lomb-Scargle periodogram to the light curves of each of 458
actually or potentially detected sources in each of four energy bands
(1.5--3 keV, 3-5 keV, 5-12 keV, and 1.5--12~keV).  A whitening
technique was applied to the periodograms before evaluation of the
statistical significance of the powers.  We discuss individual
detections with focus on relatively new findings.

\end{abstract}


\section{Introduction}
Periodic modulation of the X-ray intensity has been observed in many
X-ray binary systems (see a general discussion in White, Nagase, \&
Parmar 1995).  The periods of the detected signals range from a few
milliseconds~\citep{wvk98} to hundreds of days~\citep{hugo00}, and
are, in general, associated with spin of a neutron star, binary
orbital motion, or precession of a tilted accretion disk or other
phenomenon in which the period usually exceeds the orbital period.  In
addition, very long quasi-periods have been found to be associated
with the outbursts of X-ray transients.

The periodicities attributable to anisotropic emission from rotating
neutron stars are found mostly in high-mass X-ray binaries (HMXBs) and
in a few low-mass X-ray binaries (LMXBs).  The pulse periods range
from a few milliseconds to at least tens of minutes, and possibly
hours (e.g., Hall et al. 2000).

The known orbital periods of LMXBs range from 0.19 h to 398 h, with
the majority between 1 h and 2 d \citep{white95}. The observed
properties of the orbital modulation depend on the viewing angle.  At
low orbital inclination angles ($< 70^\circ$), X-ray orbital periods
are rarely observed.  At intermediate inclinations, periodic dipping
behavior, and in a few cases, brief eclipses of the X-rays by the
companion may be observed.  The dips may be caused by obscuration of
materials splashed above the disk plane when the gas stream from the
companion hits the accretion disk.  For high inclination systems
($>80^\circ$), X-ray eclipses have been observed in several systems
but the number is smaller than what might be expected if the systems
simply consist of a dwarf companion overflowing its Roche lobe and
transferring material to a compact object via a thin accretion
disk~\citep{joss79}.  To resolve this discrepancy, it has been
proposed~\citep{milgrom78} that the central X-ray source is hidden
behind a thick accretion disk rim while the X-rays are scattered via a
photo-ionized corona above the disk and can still be seen. The source
thus appears extended and may be partially eclipsed periodically by
the bulge at the disk rim formed by the impact of the accretion flow
on the disk.

The known orbital periods of HMXBs range from a few hours to hundreds
of days, with the majority above 1 day.  All but one of the identified
supergiant systems have nearly circular orbits with periods less than
15 d, whereas most Be star systems have eccentric orbits and periods
of several tens or hundreds of days \citep{white95}.  The orbital
periods are evident by means of several effects including eclipse of
the X-rays by the companion, phase-dependent absorption and scattering
in a stellar wind from the companion, and absorption dips caused by
non-axisymmetric structure of an accretion disk, accretion streams, or
other structures.  In Be star systems, the orbital periods may be
manifest as periodic X-ray outbursts that occur when the accretion
rate onto the compact object is enhanced around periastron passage.

Superorbital periods have been observed in both HMXBs and LMXBs.  In a
handful of cases, i.e., the on-off cycles of Her X-1 (35 d period),
LMC X-4 (30 d), and SMC X-1 ($\sim55$ d), and the 164 d jet precession
cycle of SS 433, the modulation is very likely due to the presence of
a precessing tilted accretion disk that periodically occults the X-ray
source or guides the directions of the emitted jets
\citep[e.g.,][]{katz73,levine82,pringle96}. The disk precession can be
sufficiently regular to produce quasi-coherent or relatively narrow
features in a Fourier spectrum made from a few years of data.  In
other sources, the causes of the modulation corresponding to reported
superorbital periods are not well-established.  Indeed, the
modulations may be weak and/or long compared with the duration of the
observations, so the stability and significance of the effects are
difficult to quantify.

Very long-term quasi-periodicities have been seen in several recurrent
X-ray transients. The X-ray light curves in these transients are
characterized by prominent outbursts separated by long periods of
quiescence.  The physical mechanism underlying these outbursts is
still unclear.  The favored scenario at present is some variation of
the disk instability model \citep[see the review by][]{lasota01}.
Transient behavior as well as quasi-periodicities of outburst
occurrence times are known to occur in both neutron star systems and
systems comprising black hole candidates.  Among them, the Rapid
Burster (X1730$-$333), a neutron star system, is known to exhibit
outbursts that sometimes recur every 200 d or so (but at other times
the recurrence interval has been as short as $\sim100$ d), and the
black-hole candidate 4U 1630$-$47 is known to exhibit outbursts at 500
to 700 day intervals.  A summary of detections of transient X-ray
sources observed with the \RXTE/ASM prior to the year 2000 can be
found in \cite{bradt00}.

Most of the known periodicities have been found in studies of one or a
few sources, but there have also been a few more systematic searches
for periodic signals.  \citet{priedhorsky83} analyzed data obtained
between 1969 and 1976 by the Vela 5B satellite on 4 sources in the
Centaurus region and confirmed the 41.5 day period of GX301$-$2,
strengthened the evidence for a 187 d period in 4U1145$-$619, and
reported a period of 132.5 days in GX304$-$1.  Another search using
Vela~5B data focused on 9 galactic X-ray sources in the
Aquila-Serpens-Scutum region and uncovered evidence for a 199 day
period in X1916$-$053, and a 41.6 day period in X1907+09 as well as
a 122--125 day cycle in the outbursts of the recurrent transient Aql
X-1 \citep{priedhorsky84a}.  \citet{smale92} searched for
periodicities with periods longer than 1 day in 17 confirmed or
suspected low-mass X-ray binaries also using data from Vela 5B.  They
confirmed the $\sim175$ day period in X$1820-303$
\citep{priedhorsky84b} and found evidence for a $\sim 77$ day period
in Cyg X-2, but reported that they did not convincingly detect the 199
day period in X1916$-$053 nor were long-term periods detected in 13
of the other systems. They found evidence of a 333 d period in Cyg~X-3
which could also be explained by non-periodic variability \citep[see
also][]{pt86}.  \citet{smale92} concluded that long-term cyclic
variability is rare in LMXBs. Another systematic periodicity search
was done by \citet{priedhorsky95} on 8 bright X-ray binaries detected
by WATCH/Eureca, but they did not find any previously unreported
periodicities.

Herein we report on a search for periodicities in data obtained with
the All-Sky Monitor (ASM) on the \rxte (\RXTE).  Since commencing
regular observations in 1996 March, the ASM has monitored over 400
X-ray sources.  Among these, about 150 X-ray sources have been
detected with an average intensity above 10 mCrab for at least one
week.  We have conducted a global search for evidence of periodic
behavior in 458 sources using results from the ASM produced over its
first 8.5 years of operation.

This paper is organized as follows.  Section~\ref{data} describes the
data and observational constraints.  Section~\ref{tech} describes
technical details of the Lomb-Scargle (L-S)
periodogram. Section~\ref{strategy} describes the implemented search
strategies.  Section~\ref{results} gives a summary of results and
brief discussions of selected detections.

\section{Data}\label{data}

The ASM on board \RXTE\ has been monitoring the sky routinely since
1996 March and the light curves analyzed in this paper cover roughly
the first 8.5 years (up to 2004 August 26). Normally, during each
orbit, up to $\sim$80\% of the sky is surveyed to a depth of 20 to 100
mCrab depending on the number and strength of sources in the field of
view.  A source is observed roughly 10 times a day.

The observations are analyzed to produce light curves for each of a
set of sources listed in a specific catalog.  Briefly, a set of four
linear least squares fits to the shadow patterns cast by X-rays from
sources with known positions over a $90$-s observation by one of the
three Scanning Shadow Cameras (SSCs) of the ASM yields the source
intensities in four energy bands (1.5--3, 3--5, 5--12 keV, and 1.5--12
keV).  The intensities are usually given in units of the count rates
expected if the sources were at the center of the field of view in one
of the cameras (SSC 1) in 1996 March; in these units, the $1.5$--$12$
keV Crab Nebula flux corresponds to about $75$ ASM cts s$^{-1}$.  A
detailed description of the ASM and the light curves can be found in
\citet{levine96}.

The results of the analysis are sorted by source to produce light
curves for each of the four energy bands.  Only those measurements
which satisfy a set of filtering criteria are transferred to the light
curve files; the criteria include the value of the goodness of fit
parameter (reduced $\chi^2$), and the strength of the background in
the 1.5--3 keV band.  In the standard production analysis at MIT,
weighted averages of the filtered results for each of the four energy
bands are formed in 1-d time bins in order to reduce the effects of
measurement errors on 1 day and longer time scales.

In the present analysis, we have used the source intensities estimated
on the basis of individual 90-s observations as well as 1-d
averages derived at MIT by the {\it RXTE}/ASM team.  We approximately
adjust the time of each 90-s observation of a given source for the
projected distance of the spacecraft from the barycenter of the Solar
System.

For a number of reasons, the observations of a source with the ASM are
obtained unevenly in time. The observation planning is done for 24
hour time intervals with a goal of maximizing the number of 90-s
observations consistent with a large set of constraints.  In
particular, the pointing of the spacecraft, the position of the
spacecraft in its orbit, the constraints imposed by the limited
rotation range of the ASM Drive Assembly, and the slew rate of the
Drive Assembly are among the factors that are involved in the planning
procedures.  Observations are only performed while the spacecraft
attitude is fixed and when constraints on the aspect of the SSCs
relative to the directions to the Sun and the center of the Earth are
satisfied.  The times of observations of most sources are correlated
with orbital phase because of Earth blockage and because the detector
high voltage is routinely turned off whenever the spacecraft passes
through the South Atlantic Anomaly or other regions in which high
fluxes of energetic particles are frequently encountered.  High
background levels can also be produced when a camera views the sunlit
Earth; in this case backscattered solar X-rays can be a problem when
the Sun is particularly active.  There are also other not well
understood instances of high background that appear to be increasing
with frequency and strength as the \RXTE\ mission proceeds.  For
sources near the ecliptic plane, there is an interval of typically
$\sim 30$ days each year when the Sun passes nearby as seen from the
Earth in which few useful results are obtained.  This is not only
because the ASM has to be turned off whenever the Sun is in the field
of view, but also because solar X-rays scatter off internal collimator
surfaces when the Sun is close to the field of view and produce high
and spatially non-uniform background levels, particularly below 3 keV.

In addition to the effects which tend to yield strong components in
the ``window function'' with periods of $\sim 96$ minutes (1
spacecraft orbit), 1 day, and 1 year, the orbital plane
of the satellite undergoes retrograde precession with a period
of $\sim 53$ days due to perturbations from the Earth's oblateness.
This naturally affects the pattern of observations of every source.


\section{Analysis Techniques}
\label{tech}

Periodicities in the ASM light curves have been sought by means of
Lomb-Scargle (L-S) periodograms. The L-S periodogram
\citep{lomb76,scargle82} has some advantages in the estimation of the
power density spectrum of unevenly spaced data over the classic
periodogram based on the Fast Fourier Transform (FFT).  In the L-S
periodogram, a maximum in the power occurs at the frequency which
gives the least squares fit of a sinusoidal wave to the data.  The
powers are evaluated at a set of frequencies chosen to be the same as
that for the FFT, i.e., $\omega _n =2\pi n/T$ for $n=0,1,...,N/2$,
where $T$ is the total length of the time interval covered by the data
and $N$ is the number of data points (but may be set to a different
value, see below).  In our search, we oversample by a factor of 4 so
that the frequencies are spaced by $\pi/2T$.  The goal is to ensure
the detection of a peak for a signal that is of border-line
statistical significance and to best locate the peak.  We have used
the C codes of \citet{press92} to compute the periodograms.  When we
apply the codes to the 90-s time resolution data sets, $N$ is
usually set to give a high frequency limit of 30 cycles per day rather
than the number of data points.  When we apply the codes to the 1-d
intensity averages, $N$ is naturally limited to give a maximum
frequency of one half cycle per day.  

For pure Gaussian noise, the powers calculated from the L-S
periodogram are known to be exponentially distributed.  In that case,
the probability of detecting a power $P>P_0$ among a set of $N$ powers
is
\begin{equation}
\mbox{Prob}(P>P_0)=1-(1-e^{-P_0})^N.
\end{equation}  

Two major issues must be confronted when searching for periodicities
in the L-S periodograms. First, many X-ray sources show low frequency
noise in their periodograms (most significantly at $f < 0.02$ cycles
per day).  Second, periodicities on time scales close to those noted
above, i.e, $\sim96$ m, 1 d, $\sim50$ d, 1 yr, and harmonics thereof,
are observed in many of the periodograms.  It is evident that the
patterns in the window functions (i.e., the sampling intervals)
together with the presence of strong signals in the sources
particularly at low frequencies can contribute to these observed
periodicities.  We also believe that in many cases, the commonly seen
periodicities are also due to the effects on the measured source
intensities of systematic errors whose magnitudes are correlated with
background levels and possibly other physical effects that are, in
turn, correlated with orbital phase, time of day, time of year, etc.,
in some complicated manner.  The noise level also tends to be higher
near these artificial periods.  This hinders detections of intrinsic
signals as the noise is not white.

To cope with both of these types of problems, we have adopted an
empirical method to search for and detect periodicities and to
evaluate their significance.  A periodogram can be whitened by
dividing the spectrum by an estimate of the power of the underlying
noise.  The estimate of the underlying noise power is purely
empirical; it is approximated as the local average following a method
proposed by \citet{israel96}.  In this method, the noise power at a
frequency bin $j$ is the local average of the underlying power
calculated within a fixed number of frequency bins (the bandwidth)
$\Delta j_t$.  To better evaluate the shape of the red noise at low
frequencies, the number of frequency bins to the right ($\Delta j_R$)
of bin $j$ and to the left ($\Delta j_t -\Delta j_R$) are set equal on
a logarithmic scale, i.e., $\log (j+\Delta j_R)-\log j =\log j
-\log(j-(\Delta j_t-\Delta j_R))$.  The average noise power is then
calculated in the left and right frequency bands respectively (with
bins $j-2$ to $j+2$ excluded) and an average of these two numbers
serves as the local average noise power $P_{\rm la}(j)$.  We then
compute rescaled powers $P_r(j) =P(j)/ P_{\rm la}(j)$.

For each power spectrum, the bandwidth $\Delta j_t$ is chosen in a
manner such that it maximizes the probability that the distribution of
the rescaled powers is exponential as expected for white Gaussian
noise.  To do so, we increase the bandwidth $\Delta j_t$ by factors of
2.5 from 30 frequency bins up to one tenth of the total number of
frequency bins in the spectrum and calculate, for each bin size, the
significance level that the rescaled powers follows an exponential
distribution using the Kolmogorov-Smirnov (K-S) test~\citep{press92}.
Once the optimal bandwidth is found, the spectrum is rescaled and the
rescaled powers are searched for significant signals.  In practice,
the significance level that the rescaled powers follow the exponential
distribution usually falls in the range 10\% to 90\%.  Occasionally,
the probability is very low ($<1\%$); this may be caused by the
presence of strong signals, real or artificial, their harmonics, and
beats between signals. In all cases, we assume an exponential
distribution for the rescaled powers.

Examples of both original and rescaled periodograms, in this case
based on 8.5 years of data in 90-s time bins on the source Cyg X-1,
are shown in Fig.~\ref{lomb_cygx1}.  The original periodogram is shown
in the upper panel; the known 5.6-d orbital period is apparent. In
addition, the 1-day period and its (second) harmonic, the 96 m period and
its beat with the 1-d period are clearly present. The middle panel
shows the calculated local power level and the lower panel shows the
rescaled periodogram.  The 5.6-d period is clearly detected with the
false alarm probability FAP $<0.001$.  The artifacts near 1 d$^{-1}$
and harmonics are well suppressed by the rescaling method.

For each detected periodicity, we refine our estimates of the detected
period and its uncertainty based on the following considerations.  For
a sinusoidal signal embedded in Gaussian noise, the uncertainty
in the period can be calculated theoretically when the data are evenly
sampled.  Numerical simulations have shown that the same result holds
well for unevenly sampled data (Horne \& Baliunas 1986 and references
therein).  Using the equation cited in \citet{horne86} and the relation
of the L-S power to the signal-to-noise ratio in the time domain, we
rewrite the uncertainty in the frequency in terms of the L-S power as
\begin{equation}
\delta f=\frac{3}{8} \frac{1}{T\sqrt{P_r}},
\label{P_df}
\end{equation}
where $T$ is the total length of the time interval represented by the
data set and $P_r$ is the rescaled Lomb-Scargle power.  We have
performed a simulation to confirm the validity of this estimate.  In
the simulation we add sinusoidal signals at frequencies and amplitudes
typical of our detections to all available ASM data and then use the
search procedure described above to search for periodicities within
the frequency range $0 < f < 2 f_{sim}$ where $f_{ sim}$ is the
frequency of the simulated signal.  The oversampling factor of the
periodogram is set to make the frequency bin size close to what is
expected from Eqn.~\ref{P_df}.  The measured period and its deviation
from the true value are recorded only if the signal is detected with a
FAP of $<$10\%.  We found that in $>$90\% of the cases, the deviation
between a detected period and its exact value is well bounded by the
frequency uncertainty given by Eqn.~\ref{P_df}.  Large deviations
occur only at low powers ($P_r < \sim 15$).  The threshold in our
searches is set by the FAP (see Section~\ref{strategy}) and
corresponds, in terms of power, typically to $P_r > 18$. Thus, the
estimate given in Eqn.~\ref{P_df} is a conservative upper bound for
significantly detected signals.

The periods and associated uncertainties of the detected signals are
therefore determined as follows.  We recalculate the rescaled L-S
periodogram in the frequency range $0 < f < 2 f_{init}$ where
$f_{init}$ is our initial preliminary estimate of the frequency of the
detected signal.  In this recalculation, we repeat the procedure as in
the simulation described above, that is, we use a large oversampling
factor chosen such that the frequency bin size is as close as possible
to the value estimated from Eqn.~\ref{P_df} while the period is still
detectable at 90\% confidence.  Each detected period that is listed in
column 2 of Table~\ref{period_orb}, \ref{period_long}, or
\ref{period_special} corresponds to the frequency of the bin that
yields the maximum power in the periodogram, and the quoted
uncertainty is the greater of the frequency bin size or that estimated
from Eqn.~\ref{P_df}.  For a quasi-periodicity (defined as the
frequency width $> 1/T$), we also include one half of the frequency
extent (FWHM) of the (often multiple) signal powers.

\section{Detection Strategy and Identification of Artifacts} \label{strategy}

For each periodogram, the preliminary threshold to select candidate
signals was chosen to give a false-alarm probability of $<10\%$ in the
entire periodogram.

We first reject any detections which we believe to be attributable to
causes other than behavior of the cosmic source.  The few
periodicities related to artifacts discussed previously are evident in
many sources.  The 1-day period has been detected in 50 sources in the
frequency range $f=1 \pm 0.1$ d$^{-1}$, the 96 min orbital period of
the satellite is detected in 73 sources at $f = 15\pm 0.1$ d$^{-1}$,
and the second harmonic thereof is detected in 14 sources at $f=30 \pm 0.1$
d$^{-1}$\citep[see also][]{fos2005}.  The beat between these two
periods at a frequency of 14 d$^{-1}$ is detected in 15 sources; this
is expected because the orbital motion of the satellite is in the same
sense as the rotation of the Earth. Periodicities related to the
motion of the Earth around the sun have been detected in 28 sources at
$f=0.0028 \pm 0.0005$ d$^{-1}$ (365 d period) and in 17 sources at
$f=0.0056 \pm 0.0005$ d$^{-1}$ (187 d period).  We therefore normally
exclude any detections in these frequency ranges.  We further exclude
the frequency range of $0.019 \pm 0.004$ d$^{-1}$ (detected in 23
sources) which is likely linked to the 53 d precession period of the
satellite orbit.  These frequency ranges have been chosen empirically.
We further exclude periods longer than one fourth of the duration of
the ASM light curve as the noise tends to be strong at these low
frequencies and the rescaling method does not work well in that range.
We do not reject detections in these frequency ranges if we know of an
independent report of the detection of a similar period based on data
from other instruments.

Low frequency detections from the sources GX13+1, GX1+4, GX17+2,
GX3+1, GX339$-$4, GX354$-$0, GX9+1, GX9+9, and GX5$-$1 are excluded as
each of them have large peaks (after our whitening effort) randomly
distributed throughout a broad frequency range corresponding to
periods of a few days to several hundreds of days~\citep[but
see][]{diet2005,corbet03}.

We select detections if the false alarm probability to detect the
signal of a given power in the entire periodogram of a source is (1)
$<=5\times 10^{-5}$, or (2) $<10^{-1}$ and we are aware of an earlier
report of variability at the same period within measurement
uncertainties based on independent means, such as in other wavelengths
or using other detectors.  Criterion (1) ensures that we shall expect
a total of $<0.1$ detections at such significance level in our
searches through all $\approx 2000$ periodograms if they are random
noise independently drawn from one distribution.  This has proved to
be reliable in practice (see next subsection).  Criterion (2) is
adopted because the probability of a spurious detection is
proportional to the number of frequency bins searched, and hence the
spurious detection of a previously known period is a factor of $\sim
10^4$ less probable than a spurious detection in a blind search.

It must be noted that the whitening procedure in the search technique
that we have adopted tends to reduce the sensitivity to broad
features, and in particular will reduce the sensitivity to a
quasiperiodic signal in proportion to the degree that the feature's
breadth exceeds that of a coherent signal.  This is due to the fact
that the whitening procedure tends to obliterate slowly varying
periodogram components as is clearly shown in Fig.~\ref{lomb_cygx1}.
Thus, our search is most sensitive to coherent signals.


\section{Results}\label{results}
The periodicities which satisfy the criteria described above are
listed in Table~\ref{period_orb} (pulse periods and orbital periods)
and in Table~\ref{period_long} (superorbital periods).  We also list
in Table~\ref{period_special} periodicities in five sources that
require further investigations.  For each periodicity, the tables give
the source name, detected period and uncertainty as determined from
the ASM data, and false alarm probability for a blind search through
the 8.5 years (1996 March--2004 August) of data. For several sources,
only the significance of a detection from a data set of shorter
temporal coverage is listed (see footnotes to the tables). For these
sources, the significance of the detection is lower when a data set of
longer coverage is used, possibly because of a slow deterioration of
the ASM data quality or possibly because the strength of the signals
has decreased.  We also note the nature of the systems and/or
mechanisms that have been proposed for the modulation. The entries are
grouped according to the possible nature of the systems, the
modulation mechanisms, and magnitudes of the periods.

We have listed a total of 51 significantly detected periodicities.
Two new periodicities not published before this work, IRAS04575$-$7537
(3.2 d), SAX~J1808.4$-$3658 (72.2 d), and four not previously reported
in refereed publications, i.e., those in IGR J19140+098 (13.55 d), IGR
J00370+6122 (15.67 d), IGR J11435-6109 (52.4 d), and XTE~J1716$-$389
(99 d) are included in the list of detections.  However, the
detections of the new periodicities in IRAS04575$-$7537 and
SAX~J1808.4$-$3658 may be manifestations of red noise rather than true
periodicities (see subsection \ref{special}).  The tables list
detections of five previously known X-ray periods that, to our
knowledge, have not previously been reported to be apparent in ASM
data. They are the orbital periods of the LMXBs EXO0748$-$676 (3.84
h), X1658$-$298 (7.11 h), X1822$-$371 (5.57 h), X1916$-$053 (3000.7 s), and a superorbital period in GRS1747$-$312
(147 d).  Our results also confirm the periodicities previously
reported on the basis of X-ray intensity variations in (often much
less) ASM data in X0114+650 (2.74 h, 11.599 d, 30.8 d), X1908+075 (4.4
d), XTE~J1855$-$026 (6.1 d), LSI+61303 (26.2 d), GRO J2058+42 (55.1
d), X2206+543 (9.6 d), X0726$-$260 (34.4 d), RX~J0812.4$-$3115 (81 d),
IGR J$11435-6109$ (52.4 d), IGR J$00370+6122$ (15.67 d), and IGR
J$19140+098$ (13.55 d).  Among them, the 30.8 d periodicity in
X0114+650 was not apparent in data sets with length shorter than 6.5
years.  We did not find significant detections of a 99 d or 189 d
period in LMC X-3, a 42~d period in X1907+097, or a 200~d period in
X1916$-$053 \citep[cf.][]{cowley91,priedhorsky84a}.  Brief discussions
of all the sources mentioned above with references to previous work
are given in sections \ref{focus1} and \ref{focus2} except for
X1822$-$371\citep{mason80,white82}, LSI+61303
(Fig.~\ref{lomb_lsi+61303}); \citep[]{paredes97}, X1907+097
\citep[and references therein]{baykal01, cox05}, and X2127+119
\citep{ilovaisky93}.


With observations spanning over 8 years, the ASM data sometimes yield
a measurement of a period which is comparable to or more accurate than
previous measurements.  These include the orbital periods of
X1916$-$053, X1624$-$490, X0114+650, SAX~J2103.5+4545 (see discussions
in section~\ref{focus1}) and X2127+119 \citep[$P_{orb}=0.713014 \pm
0.000001$ d,][]{ilovaisky93}.

We offer brief discussions of several individual detections in the
following subsection.  The ASM light curves and periodograms are shown
for some of the individual sources discussed.  Periodograms for
low-frequency signals are often shown in the original Lomb-Scargle
periodograms to show possible features around the signals. Rescaled
periodograms are shown for detections at higher frequencies.

For each periodicity that is likely to be more or less coherent over
the time interval used in the analysis for a given source, a set of
folded light curves in the 1.5--3, 3--5, 5--12, and 1.5--12 keV energy
bands has been made.  The time interval used to make the folded light
curve is generally 8.5 years, i.e., ending in 2004 August, except for
four cases as noted in Table 1.  The times of the data were corrected
to the Solar System barycenter and folded according to the periods
listed in Tables 1, 2, and 3 except for the light curve of GRO
J2058+42 which was folded at twice the period listed in Table 1.
Weights given by the inverse square of the estimated errors of the
individual measurements were used to obtain the average intensity for
each phase bin.  The folded light curves are shown in
Figures~\ref{fold_1} through~\ref{fold_5}.

\subsection{Spin and Orbital Periods}
\label{focus1}

{\bf X0114+650} (2S0114+650) is associated with a B1 Ia optical
counterpart (Reig et al. 1996).  An orbital period of 11.591$\pm$0.003
days was reported from optical radial velocity measurements
\citep{crampton85}.  Finley, Belloni, \& Cassinelli (1992) reported a
2.78 $\pm$ 0.01 h period for which there was evidence from X-ray
observations with several satellites (ROSAT, Ginga, and EXOSAT)
although the number of 2.78 h cycles in each observation was limited.
The same period was also apparent in further ROSAT observations
(Finley, Taylor, \& Belloni 1994).  \citet{corbet99a} analyzed the ASM
light curve of X0114+650 constructed from 2.5 years of observations
and found modulations at both the optically derived orbital period and
at periods close to 2.74 hours. Their analysis of subintervals of data
showed changes in the latter period which were consistent with those
expected from accretion onto a neutron star. Thus, \citet{corbet99a}
concluded that the $\sim2.7$ h period may represent the longest known
neutron star spin period.


Using the 8.5 years of ASM data on X0114+650, we have significantly
detected the 11.6 d orbital period and a 31 d period
(Fig.~\ref{lomb_x0114}).  The $\sim2.74$ h (presumable) spin period
can be detected significantly in a blind search using only $< 4.5$
years of ASM data.  We show in time-frequency diagrams
(Fig.~\ref{t_f_x0114_1}) how the frequency corresponding to the 2.74 h
period feature appears roughly every 0.5 years since 1996 March.  The
evolution of the $\sim$ 2.74 h period is apparent. If this is indeed
the neutron star spin frequency, the star must undergo dramatic spin
changes on time scales $< \sim 0.5$ year.


Detection of the 31 d period has been reported recently by
\citet{farrell04}. The origin of this periodicity is unclear.  The 31
day period is not detectable using only the first 6.5 years of the ASM
results on this source; it is most significant at the last half year
when the otherwise prominent 11-d period becomes less significant.

{\bf X0352+30} (X~Per) has been classified as a Be/neutron star
system.  The neutron star shows pulsations at a period of $\sim 837$ s
\citep{white76}.  An orbital period of 250 d was revealed in a pulse
timing analysis \citep{hugo00}.  The pulsation period cannot be
detected by simply applying our whitening technique to the ASM data
because there are variations in the pulse period due to accretion
torques that spread the power from the pulsations into many frequency
bins.  However, a signal appears in the unrescaled periodogram at a
period of 837.8 s (Fig.~\ref{lomb_xper}).  It is significant given
that the period was previously known.  We found no evidence of the 250
d orbital period in the ASM data.

{\bf EXO0748$-$676} is one of a few LMXBs in which full eclipses of
the compact object are seen.  It is also an X-ray dipping source.  The
eclipse lasts about 500 s and reoccurs at the orbital period of 3.8 h
\citep{parmar86,parmar91}.  Even though the eclipses are very brief
and the source is faint, $F_X$ is typically $\sim8$ mCrab, the orbital
period is detected at 3.8241 h with a FAP of $6\times 10^{-7}$ in the
3--5 keV energy band with the 8.5 year ASM data set
(Fig.~\ref{lomb_exo0748}). It is also detected in the 1.5--12 keV band
with lower confidence ( FAP $\sim 1\times 10^{-4}$). The periodic
modulation appears to be highly energy dependent as it is not detected
at or above 90\% confidence in the 1.5--3 keV or 5-12 keV band.

{\bf X1658$-$298}, also known as MXB1659$-$29, is one of the few known
X-ray eclipsing and dipping sources.  It is a soft X-ray transient
source that reemerged in 1999 April after 21 years of quiescence.
The 7.11 h orbital period was observed in X-rays by
\citet{cominsky89}.  The most recent report regarding its 7.11 h
orbital period in X-rays after its reemergence was made by
\citet{wachter00} using \RXTE/PCA data.  In the present search the
period is also detected significantly in the 8.5 year, 1.5--12 keV
band data set (Fig.~\ref{lomb_x1658}).  The signal is also visible in
the 3--5 keV band with a FAP $\sim 2 \times 10^{-2}$.

{\bf 4U1624$-$490} is the so-called ``big-dipper'', one of the most
unusual X-ray dipping sources.  Its 21 h orbital period
\citep{watson85} is the longest among all other dipping sources known.
The dips can be as deep as $\sim 75\%$ in the 1--10 keV band.  The ASM
observations have provided the best data at this writing to determine
its orbital period.  We provide a revised period of 0.86990(2) d which
is consistent with the value of 0.86991(13) d ($20.8778\pm0.003$ h)
based on the analysis of 4.5 years of ASM data \citep{smale01}.

{\bf X1916$-$053} is the most compact known X-ray dipping source.  Its
orbital period was reported on the basis of X-ray observations to be
in the range of 2985--3015 s \citep{walter82,
white82,smale89,yoshida93, church97}. A 3027 second optical period,
however, is found to be stable for at least 7 years
\citep{callanan95}.  We have detected a coherent $3000.645 \pm 0.004$
s orbital period using 8.5 years of the ASM data
(Fig.~\ref{lomb_x1916}).  This agrees with a recent report of
$3000.6508 \pm 0.0009$ s based on X-ray data from the \RXTE/PCA in
combination with data from the Einstein, EXOSAT, and Ginga satellites
\citep{chou01}. It has also been reported in the same paper that the
X-ray orbital period is not consistent with the $3027.5510 \pm 0.0052$
s period determined from optical observations. A hierarchical triple
system was proposed to explain the discrepancy.

{\bf X1908+075} is an HMXB with a 4.4 d orbital period that was found
using $\sim 3$ years of ASM data \citep{wen00a}.  Observations with
the \RXTE/PCA resulted in the detection of a 605 s pulse period and of
orbital Doppler delays \citep{levine04}.  Orbital phase dependent low
energy absorption led to estimates of the system inclination and to
the mass of the companion star.  \citet{morel05} have reported a
likely infrared companion that could be an OB supergiant.

{\bf Cyg~X-1} is an HMXB that is believed to comprise an accreting
$\sim 10$ solar mass black hole.  Two physically distinct states of
the X-ray source have been observed: the hard state and the soft
state.  Most of the time, Cyg X-1 stays in the hard state where its
$2$--$10$ keV luminosity is low and the energy spectrum is hard. Every
few years, Cyg X-1 undergoes a transition to the soft state and stays
there for weeks to months before returning to the hard state.  During
the transition to the soft state, the $2$--$10$ keV luminosity
increases, often by a factor of more than $4$, and the energy spectrum
becomes softer (see reviews by Oda 1977; Liang \& Nolan 1984 and
references therein). The ASM data reveal the 5.6 d orbital period in
the low state but not in the high state \citep{wen99}.  The
significance level given in Table \ref{period_orb} is based on an
analysis of the 1 d averages including both the high and the low state
data (see Fig.~\ref{lomb_cygx1}).

{\bf XTE~J1855$-$026} was serendipitously discovered with the \RXTE/PCA
detector during slews along the Galactic plane \citep{corbet99b}. The
source exhibited pulsations at a period of 361 s and the ASM light
curve revealed a period of 6.067$\pm$ 0.004 d which was interpreted
as the orbital period of the system. A subsequent extended PCA light
curve covering an entire 6.067 day period enabled a pulse timing orbit
to be obtained which gave a mass function of $\sim 16$ \msun\
\citep{corbet02}.  This mass function, together with the detection of
an extended near total eclipse, showed that the system was very likely
a binary consisting of a neutron star accreting from the wind of an
early type supergiant.  An improved position was obtained by the same
authors using ASCA data. This refined position improved the quality of
the ASM light curve and, together with the use of the full 8.5 years
of ASM data, yielded an orbital period estimate of 6.0752 $\pm$ 0.0008
days.






{\bf IGR J19140+098} (IGR J19140+0951) was discovered with INTEGRAL
(Hannikainen et al. 2004). A brief observation with the RXTE
Proportional Counter Array did not detect any pulsations (Swank \&
Markwardt 2003).  Corbet, Hannikainen \& Remillard (2004) reported the
discovery of a 13.55 day periodicity in the ASM light curve of this
source which they interpreted as the orbital period of an X-ray
binary. Corbet et al. (2004) noted that this periodicity was present
in just the first 3 years of ASM data and so the source was not a
recent transient. 

{\bf IGR J00370+6122} (RX J0037.2+6121) was discovered in a long
INTEGRAL observation of the Cassiopeia region by den Hartog et
al. (2004). A $15.665 \pm 0.006$ day period was also found in the ASM
light curve and a B supergiant counterpart was identified by den
Hartog et al.  The INTEGRAL error box contains the ROSAT source 1RXS
J003709.6+612131.  No pulsations have yet been found.  Reig et
al. (2005) suggest that IGR J00370+6122 shares some characteristics
with the extreme LMC transient A0538-66.

{\bf X2206+543} (4U2206+54) was identified with a Be star by
\citet{steiner84}.  \citet{corbet01} investigated the ASM light curve
of this source and found a 9.6 day periodicity. If this is the orbital
period, as appears likely, it is relatively short for a Be/neutron
star binary and suggests that the X-ray luminosity should be high. The
measured flux, however, corresponds to a luminosity of only $\sim
10^{35}$ ergs s$^{-1}$ if the source is at the estimated distance of
$\sim 3$ kpc.  PCA observations and archival {\it EXOSAT} observations
have failed to reveal any evidence of pulsations \citep{corbet01} such
as reported earlier by \citet{saras92}.  The short orbital period and
low luminosity may be at least partially reconciled by the
reclassification of the optical counterpart as a peculiar O9 III-V
star \citep{neg01}.

{\bf SAX~J2103.5+4545} is an X-ray pulsar with a 358 s pulse period in
an orbit with a $12.68 \pm 0.25$ d period \citep{hull98,baykal00}.
\citet{reig04} identified the optical counterpart as a Be star.  The
orbital period is easily detectable in the ASM data, and our analysis
yields $P_{orb} = 12.673 \pm 0.004$ d (see
Fig.~\ref{lomb_saxj2103}). The orbital period is one of the shortest
known for a Be/neutron star system.

{\bf X0726$-$260} (4U0728$-$25) has an optical counterpart of spectral
type O8-9Ve \citep{neg96}.  From an analysis of an \RXTE/ASM light
curve spanning 1.5 years, \citet{corbet97} found evidence for a 34.5
day orbital period.  This was apparently confirmed when a PCA
observation made at the predicted time of flux maximum showed the
source to be in a bright state.  \citet{corbet97} also found 103.2 s
pulsations in the X-ray flux.  We have detected the 34.5 d period in
the 8.5 year data set with improved significance.

{\bf IGR J11435$-$6109} was reported as a hard X-ray transient by
Grebenev et al. (2004). Evidence of the possible presence of 166~s
pulsations was reported by Swank \& Markwardt (2004) which was
confirmed by BeppoSAX Wide Field Camera observations by in 't Zand \&
Heise (2004).  In 't Zand \& Heise also reported a possible 52.5 day
outburst recurrence period from detections in 1996-1997 and 2001-2002,
but non-detection during the intervening period. This outburst period
was confirmed by Corbet \& Remillard (2005) from RXTE/ASM
observations.  An initial suggestion that IGR J11435$-$6109 corresponds
to 1RXS J114358.1$-$610736 is apparently excluded by optical
observations of the optical counterpart of 1RXS J114358.1$-$61073 by
Torrejon \& Negueruela (2004) who instead propose a Be star
counterpart which is 1.2 arc minutes away from 2E 1141.6$-$6050.

{\bf GRO~J2058+42} is a transient 198 s X-ray pulsar suspected also to
be a Be-star system \citep{wilson98}.  Both ASM and BATSE data show a
number of outbursts from this source at intervals of 55 days, although
a set of outbursts that appeared at intervals of 110 days were
particularly strong in the BATSE data \citep{crp97,wilson98}.  It is
not clear which of these intervals, if either, is the orbital period.
The detection of the 55 d period is unambiguous in the ASM data even
though it nearly coincides with the precession period of the orbital
plane of the \RXTE\ satellite.
 
{\bf RX~J0812.4$-$3114} was found to exhibit pulsations at a period of
31.9 s in data obtained with the \RXTE/PCA by \citet{reig99} who
concluded that it is a Be/neutron star binary system.  As seen in the
\RXTE/ASM light curve, it underwent a transition in early 1998 from an
inactive state to a state wherein short outbursts occurred every $\sim
80$ days. \citet{corbet00} have interpreted the interval between
outbursts as the orbital period.  Our visual inspection of the ASM
light curve indicates that the series of outbursts lasted for
$\sim800$ days, and our analysis confirms the detection of the
$\sim80$ d period.


\subsection{Superorbital Periods}
\label{focus2}

{\bf XTE~J1716$-$389} is probably the same source as KS1716$-$389,
which was seen in 1994 with the TTM instrument
\citep{aleksandrovich95}.  A $\sim99$ d period is detected in the 8.5
year ASM data set (Fig.~\ref{lomb_x1716}). However, our most
significant detection was made using an ASM data set spanning a 4.5
year interval beginning in 1996 February in which a $97.51 \pm 0.12$ d
period is detected in the Lomb-Scargle periodogram with a false alarm
probability less than $3\times 10^{-7}$.  The peak in the periodogram
is narrow with a width (FWHM) of 5 days indicating detection of a
highly periodic modulation.  The uncertainty in the period is the
estimated standard deviation assuming a single sinusoidal signal with
Gaussian noise~\citep{horne86}. No other periodicities, excluding
known artifacts, have been detected in the ASM data in the frequency
range 0.001 to 30 d$^{-1}$.

The ASM light curve shows the source to have generally persisted at an
intensity of $\sim$25 mCrab from the beginning of operation of the ASM
in 1996 February until approximately 2003 June 10 (MJD 52800) after
which the intensity is less than 10 mCrab. Despite its relatively low
intensity, prominent dips, which may be associated with the
$\sim$100-d period, are apparent in the light curve
(Fig.~\ref{ltc_x1716_asm}).  In the dips the intensity may decrease to
less than $10\%$ of the long-term average count rate.  The total
duration of a dip may be as much as $\sim$40 days.  Results from both
ASM and PCA observations illustrating the variation of the intensity
through three dips are shown in Fig.~\ref{ltc_x1716_asm_pca}.  An
analysis of the timing and spectral properties of the source may be
found in \citet{wen01c}.  Throughout the broad dips, the PCA spectra
exhibit variable but enhanced absorption.  The ephemeris for the
intensity minima based on the data for the interval 1996
February--2000 August is:
\begin{equation}
T_n = {\rm MJD}\ 50174.24\pm 0.72+n\times (97.51 \pm 0.12).
\end{equation}
These times of minima were obtained from a fit of a sinusoid
to the ASM data using the Levenberg-Marquardt (L-M)
method~\citep{press92} which, in turn, relies on $\chi^2$
minimization.  The uncertainty in $T_n$ is the
estimated standard deviation from the fit.

If the X-ray emission from XTE~J1716$-$389 is powered by Roche lobe
overflow in a circular or nearly circular binary orbit with a 98 day
period, then the mass donor star would need to be in the giant phase
of its evolution \citep[see, e.g.,][]{verbunt95}.  If, alternatively,
the mass transfer is the result of accretion from a wind in an HMXB,
the companion star would likely need to be fairly massive and
evolved so that sufficient mass to power the X-ray source would be
captured by the compact object.  Precession of a tilted disk like that
observed in Her~X-1, LMC~X-4, or SMC~X-1 is perhaps a more viable
model in which case the orbital period would be substantially shorter
than 98 days.

{\bf SS~433} is a well-known galactic jet source.  Optical emission
lines that undergo dramatic motions in wavelength show that the system
emits two oppositely directed jets at velocities of $\sim 0.26$c  which
precess with a period of $\sim164$ days \citep[see,
e.g.,][]{margon84,eiken01}.  The jets are also manifest in many other
ways, notably including radio emission \citep[e.g.,][]{blundell01} and
the Doppler shifts of X-ray emission lines
\citep[e.g.,][]{kotani96,mcs02,mig02}.  The system is believed to be a
binary comprising a massive normal-type star and an accreting compact
object; thus it may be very similar to an HMXB.  The orbital period is
13.1 days \citep{crampton80}.  The nature of the engine that produces
the jet and of the compact object itself are still unknown.

We have detected at a high degree of confidence a period of 162 d in the
ASM data. This period is consistent with the previously established
precession period (Fig.~\ref{lomb_ss433}).  A peak is also found in
the periodogram at $13.090 \pm 0.001$ d, which is consistent with the
orbital period.  In the rescaled power spectrum, the false alarm
probability of finding such a peak at a given frequency is 0.005.
Evidence for detection of these periodicities in the ASM data has
previously been presented by \citet{gies02}.

{{\bf X1820$-$303} resides in the globular cluster NGC~6624.  It was
the first identified Type I X-ray burster \citep{grindlay76}. The
intensity of the source is known to vary with a period of about 176
days \citep{priedhorsky84b,smale92}.  This periodicity is easily seen
in the ASM light curve and is clearly detected in the rescaled L-S
periodogram (Fig.~\ref{lomb_x1820}).  We did not detect the known 685
s orbital modulation \citep{stella87} in an extended search.

{\bf GRS1747$-$312} is a bright transient X-ray source in the globular
cluster Terzan 6.  It exhibits complete eclipses that recur at the
orbital period of 0.52 d \citep{zand00,zand03}.  Outbursts occur
approximately every 4.5 months and type I thermonuclear X-ray bursts
have been found during some of them.  We have detected a 147 d period
consistent with previous observations of these quasi-periodic
outbursts (Fig.~\ref{lomb_grs1747}).  The orbital period is not
detected.  



{\bf SMC X-1} is an eclipsing HMXB/NS system with a 3.89 d X-ray
orbital period \citep{schreier72,levine93}.  A $\sim 60$ d
quasi-periodicity was suggested by \citet{gruber84} from {\it HEAO} 1
(A4) data, and was confirmed by {\it CGRO}/BATSE observations and
early \RXTE/ASM results \citep{wojdowski98,clarkson03}.  In our
search, there are a number of prominent features in the periodogram
corresponding to periods around $\sim 56$ d (Fig.~\ref{lomb_smcx1}).
In particular, peaks which are $\gtrsim 0.5$ as large as the most
prominent peak occur in the period range $\sim 53$ to $\sim 59$ days.
The existence of closely spaced multiple peaks suggest that this is
inherently a broad feature. This is consistent with previous
conclusions that the $\sim56$ day cycle is quasi-periodic.  The 3.9 d
orbital period and its harmonics are detected significantly.  As
expected, they do not show structure similar to that seen around the
$\sim56$ d feature.


{\bf Cyg~X-2} is one of the brightest LMXB/NS systems.  It has an
orbital period of 9.8444 d \citep{cowley79,casares98}.
\citet{smale92} reported the presence of a 77~d period in the X-ray
intensity.  \citet{paul00}, on the basis of the first few years of
\RXTE/ASM data, concluded that there is no particular stable long-term
period.  However, \citet{boyd04} report that the long-term variations
in Cyg X-2 can be attributed, at least in part, to about half a dozen
periodicities with periods that are integer multiples of the orbital
period.  In agreement with these results, we find no single dominating
periodicity in the 8.5 year data set. Instead, multiple peaks are
apparent in the L-S periodogram
(Fig.~\ref{lomb_cygx2}). Quasi-periodic intensity variations are
evident in the ASM light curve (Fig.~\ref{ltc_cygx2}).


{\bf LMC~X-3} is one of the persistent X-ray emitting stellar mass
black hole candidates.  Multiple peaks are visible in its periodogram at
time scales in the range 100--500 d (Fig.~\ref{lomb_lmcx3}).  The
variability is clearly apparent in the ASM light curves
(Fig.~\ref{ltc_lmcx3}).  Our results are consistent with a previous
report by \cite{paul00} that there is no stable long period in this
source. Neither the 99 d nor 189 d period \citep{cowley91} is
significant  in the rescaled periodogram.

\subsection{Special Cases}
\label{special}

In this subsection, we discuss our results for five sources that
formally indicate the presence of periodicities but which should be
regarded as tentative and thus in need of confirmation in independent
investigations.  Two types of possible periodicities that showed up in
the course of our search are discussed here, (1) previously unknown
periodicities detected with marginal significance by our method (3.23
d in IRAS04575$-$7537, 72 d in SAX~J1808.4$-$3658), and (2)
previously known periodicities that are apparent as broad and weak
features in the original Lomb-Scargle periodograms and therefore
failed the significance test (X0115+634, X1942+274, 4U1145$-$619).
We note that the two type (1) candidate periodicities only met our
detection criteria in the analysis of the periodograms made from 1-d
average light curves in which the threshold in terms of power
(although same in FAP) is somewhat lower than that used in the search
of the higher time resolution data.

{\bf IRAS04575$-$7537} is a Seyfert 2 galaxy at redshift of 0.018
\citep{polletta96} and galactic latitude $b=-33 \arcdeg$.  The
detection of a 3.23 d X-ray period is based upon the periodogram shown
in Fig.~\ref{lomb_iras04575}.  The period is detected at marginal
significance in a periodogram made using data in 1-d time bins for the
1.2--15 keV band; a peak is also visible in the periodogram for the
5--12 keV band and in a periodogram made directly using 90-s
intensities but is not sufficiently high to meet our formal detection
criterion.  A phase-averaged light curve made by folding at this
period indicates a smooth modulation (see Fig.~\ref{fold_4}). Since
the significance of the periodogram peak is marginal, we cannot fully
exclude the possibility that it is due to a statistical fluctuation.
If it is a real periodicity, it would need to result, e.g., from a
foreground high galactic latitude X-ray binary.

{\bf X0115+634} (4U0115+634) has been long suspected to be a Be star
system and it is known to comprise a 3.6~s pulsar \citep{cominsky78}.
Its optical counterpart was recently reclassified an O9e star
\citep{unger98}.  Within the 8.5 years covered by the ASM
observations, it underwent two major outbursts and some ``failed''
ones on intervals of $\sim 500$ d. We found that the the known 24~d
orbital period \citep{rappaport78} appears to be the highest peak in
the Lomb-Scargle periodogram for the first 3 years of ASM data (before
the first outbursts, $<$ MJD 51142) (Fig.~\ref{x0115}).  However,
there are broad features around the period.  No significant detection
was made in the rescaled periodogram.

{\bf SAX~J1808.4$-$3658} was found by \citet{wvk98} to exhibit
pulsations at a frequency of 401 Hz which made it the first discovered
accreting millisecond X-ray pulsar. A pulse timing analysis showed
Doppler shifts that reveal the orbital period of about 7249~s
\citep{chakrabarty98}.  In the present analysis, the periodograms of
the 3--5 keV and 1.5--3 keV band data for all 8.5 years averaged in
1-d time bins show a peak at a 72 d period (Fig.~\ref{lomb_saxj1808}).
The 72 d period also appears as the location of the maximum power in
the periodograms made using 90-s time bins in the 1.5--3 keV, 3--5
keV, and 1.5--12 keV bands, but in these cases the strengths are not
significant in terms of our detection criteria.  The folded light
curve (Fig.~\ref{fold_4}) appears to indicate that this detection is
highly significant.  However, this is misleading because the
statistics are dominated by the low frequency behavior of the source
which is not reflected in the error bars.  We find no evidence of
intensity variation at the orbital period.

{\bf X1942+274} (GRO~J1944+26) was first discovered as a 15.8 s pulsar
\citep{fishman89}.  The \RXTE/ASM light curve of this source shows
outbursts between 1998 and 2001 that recur more or less every $\sim
80$ days, and that may be characteristic of a Be/NS system.  There was
a previous report of a $\sim 80$ d period of the outbursts
\citep{campana99}.  However, the orbital period is 169.2 d
\citep{wfcn2003}.  The outbursts occur roughly twice per orbit at
phases that are not stable.  The $\sim80$-d (quasi-)periodicity is
evident in the ASM light curves of X1942+274 (Fig.~\ref{ltc_x1942}).
Such periodicity also appears as the highest peak in the Lomb-Scargle
periodogram in the 5--12 keV band (Fig.~\ref{power_x1942}) as well as
1.5--12 keV band. However, in the rescaled periodogram, the
periodicity is not significant due to the appearance of the broad
features around this period.  

{\bf 4U1145$-$619} is a persistent but highly variable Be/NS system.
Analysis of the long-term X-ray behavior using 4 years of Ariel V Sky
Survey Instrument (SSI) observations \citep{watson81} showed four
outbursts at approximately 6 month intervals.  The outbursts are
typically of 10-d duration with flux levels increasing by a factor of
$\sim 5$.  The seven years data from Vela 5B of this source is
dominated by one strong outburst. In their paper, the power spectrum
for data excluding the burst showed no sign of the 187.5 d
period. However, the epoch-folding analysis showed some evidence for
half this period \citep{priedhorsky83}. Three separate EXOSAT
observations \citep{warwick85} confirmed the intensity rise and
decrease that are consistent with the ephemeris of the outbursts from
previous observations.

With its 9 years' baseline combined with reasonable sensitivity and
excellent coverage, the ASM data provide the best long-term
observations for this source. The ASM light curve and hardness ratios
are shown in Fig.~\ref{ltc_hr_x1145}.  At least the first three
outbursts are visible in the ASM data. The times of these outbursts
are consistent with the ephemeris given from previous observations
\citep{priedhorsky83}. Four other possible outbursts are also
consistent with the ephemeris (numbers 8, 12, 13, 14 from the left).
The strength of the outbursts seems to be highly variable from cycle
to cycle, consistent with the Vela 5B observations.

The Lomb-Scargle periodogram of this source for 8.5 years of ASM light
curves is shown in Fig.~\ref{x1145_lomb_390}.  A peak at $187.0\pm
3.2$ d is detected clearly. The uncertainty in the period was
estimated according to the frequency bin size.  Side lobes indicating
modulation on a timescale of $\sim3.3$ yrs are visible.

\section{Summary}

We have presented results from a systematic periodicity search through
the first 8.5 years of source intensity measurements made using the
\RXTE/ASM.  The ASM data base is an incredibly valuable resource that
has allowed us to carry out this search in a uniform fashion using the
intensity records for a large number of X-ray sources.  Our search has
served the purposes of discovering and measuring new periodicities and
providing a means for improving the level of knowledge of previously
known periodic phenomena in X-ray sources.  The detection strategy we
have adopted is conservative in general.  It, hopefully, can be useful
in clarifying the authenticity of some previously claimed
periodicities.

The results of our search demonstrate that orbital modulation is more
readily detected in HMXBs than in LMXBs \citep{joss79}. A total of 33
orbital periods have been securely detected in our search.  Only eight
of them are believed to be in LMXBs, while twenty-four are in HMXBs
with sixteen in supergiant systems and eight in Be-star systems.  Out
of the $\sim$ 400 X-ray sources monitored with the ASM, over 100 of
them are believed to be LMXBs and only $<60$ are believed to be HMXBs.
Yet it is apparent from the present results, that there are far fewer
orbital periodicities detected in the LMXBs than in the HMXBs.  In
particular, the fraction of eclipses observed in LMXBs is $<3\%$, well
below that of the HMXB systems, consistent with previous observations
that the number of eclipses detected in LMXBs is well below the
expected rate if the companion stars are dwarf stars that fill their
Roche lobe and if the accretion disks are thin.

\acknowledgements

We gratefully acknowledge Gianluca Israel, Andrew Peele, Saul
Rappaport, Ronald Remillard, and Jean Swank for helpful contributions
and conversations, the efforts of the RXTE/ASM groups at MIT and
NASA/GSFC, and the RXTE mission support groups at GSFC. We are
grateful to an anonymous referee for several helpful comments.  We
acknowledge the support of NASA through Contract NAS 5-30612 (MIT).


\pagebreak

\plotone{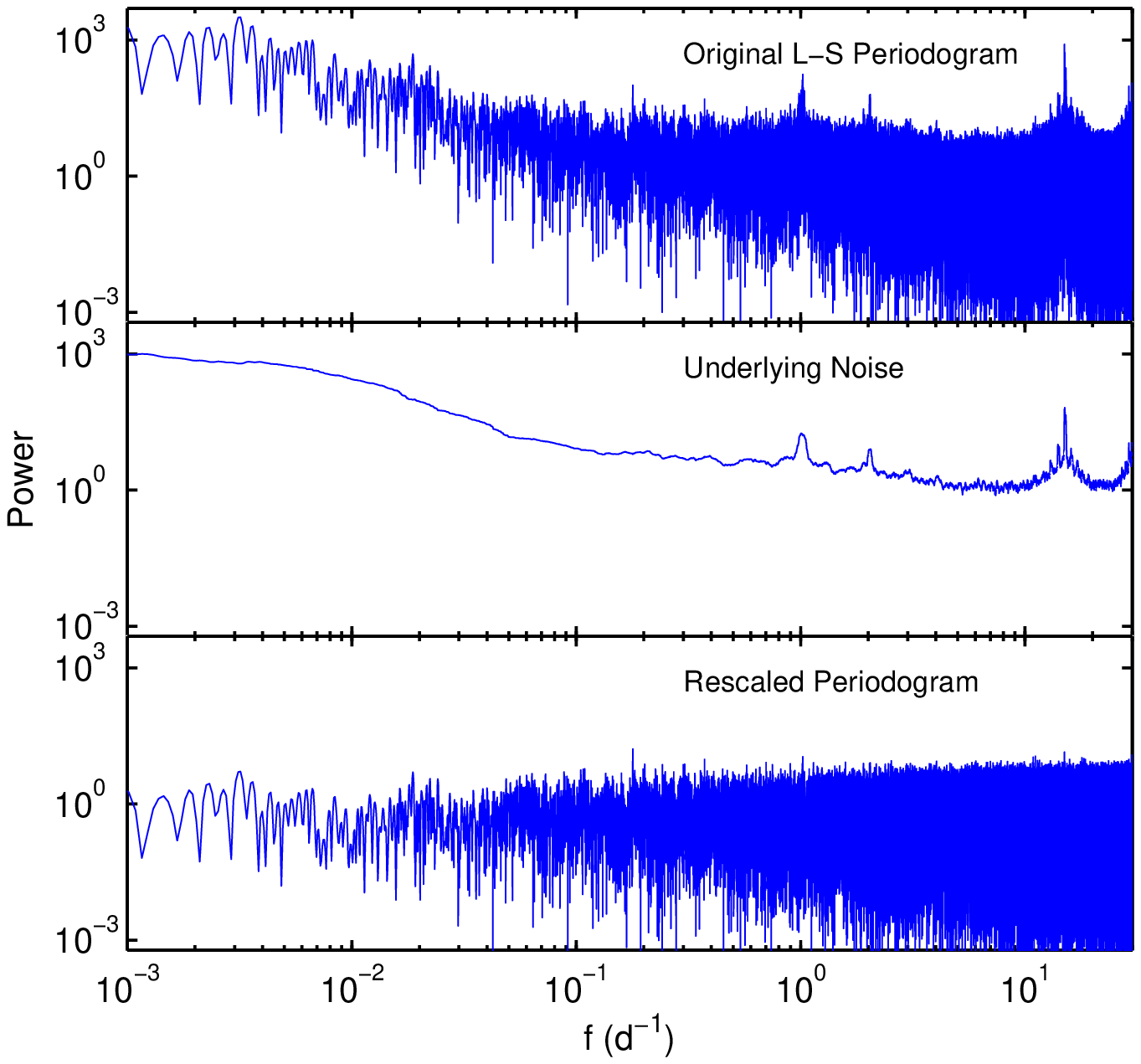}
\begin{figure}
\caption{Lomb-Scargle periodogram of Cyg~X-1, the calculated
 underlying noise, and the rescaled periodogram (see
 text). The 5.6 d orbital period of Cyg~X-1 is the highest
 power in the rescaled periodogram and is detected at a 99.99
 \% significance level.}\label{lomb_cygx1}
\end{figure}

\plotone{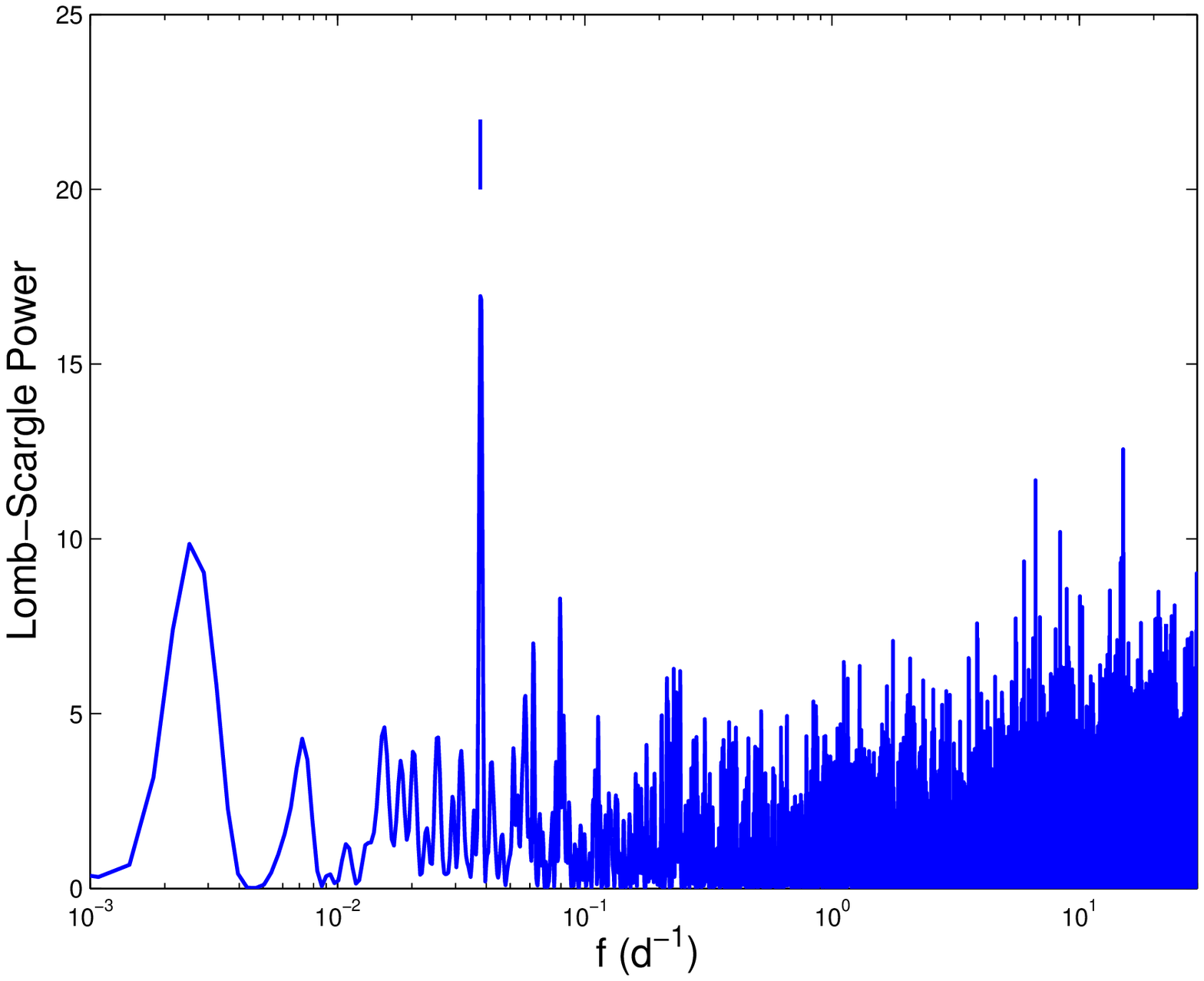}
\begin{figure}
\caption{Lomb-Scargle periodogram of the first 2.5 years of 90-s time
resolution ASM data on LSI+61303. The 26.48 d period is indicated
with a short vertical line.}\label{lomb_lsi+61303}
\end{figure}

\begin{figure}
\includegraphics[clip, width=5.5in]{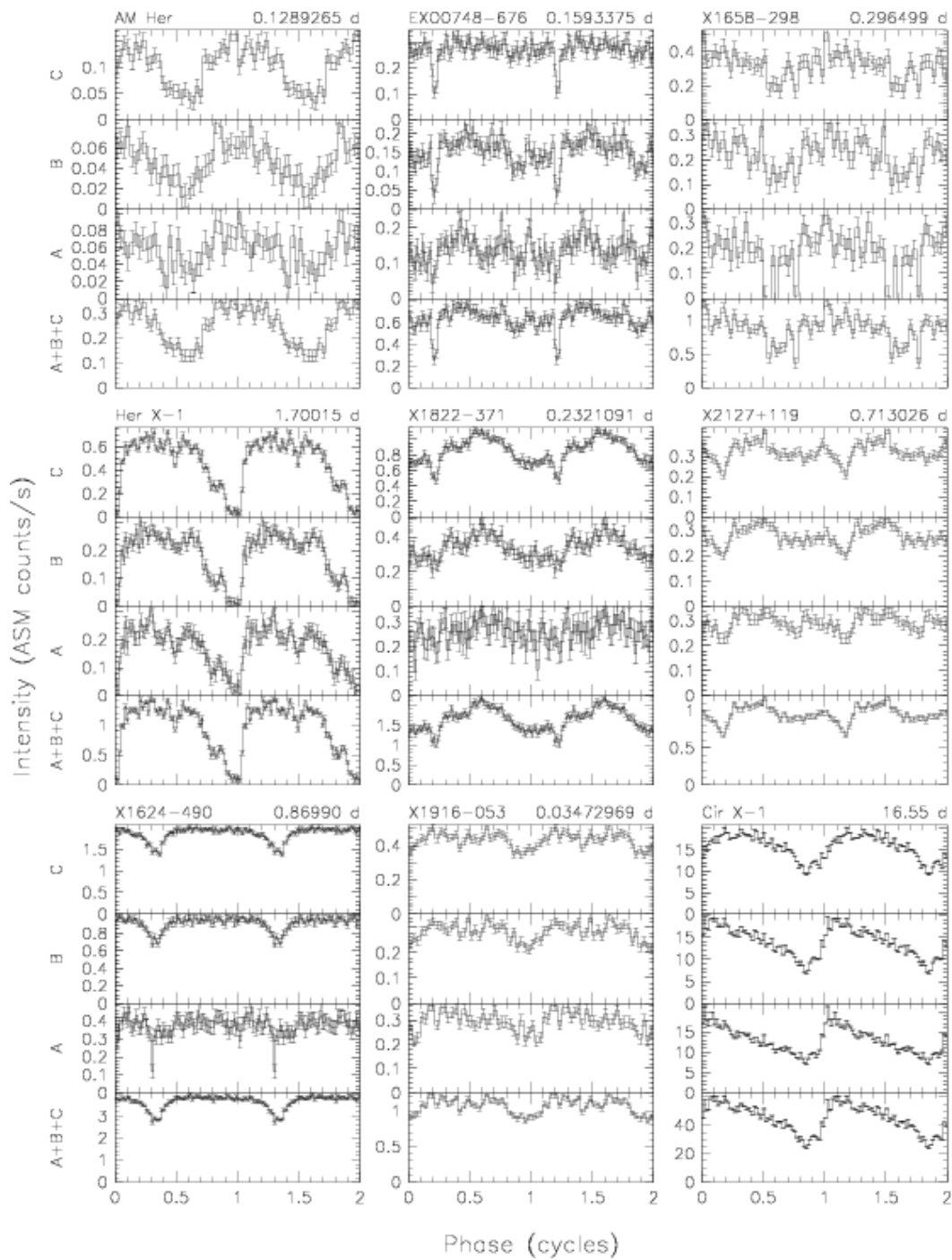}
\caption{Light curves for one polar (AM Her) and 8 LMXBs folded at the
orbital periods from Table 1.  For each source, the folded light curve
is given for the 1.5--3 keV (A), 3--5 keV (B), 5--12 keV (C), and
1.5--12 keV (A+B+C) bands. For clarity, the
folded light curve is shown for two full periods.}\label{fold_1}
\end{figure}

\begin{figure}
\includegraphics[clip, width=5.5in]{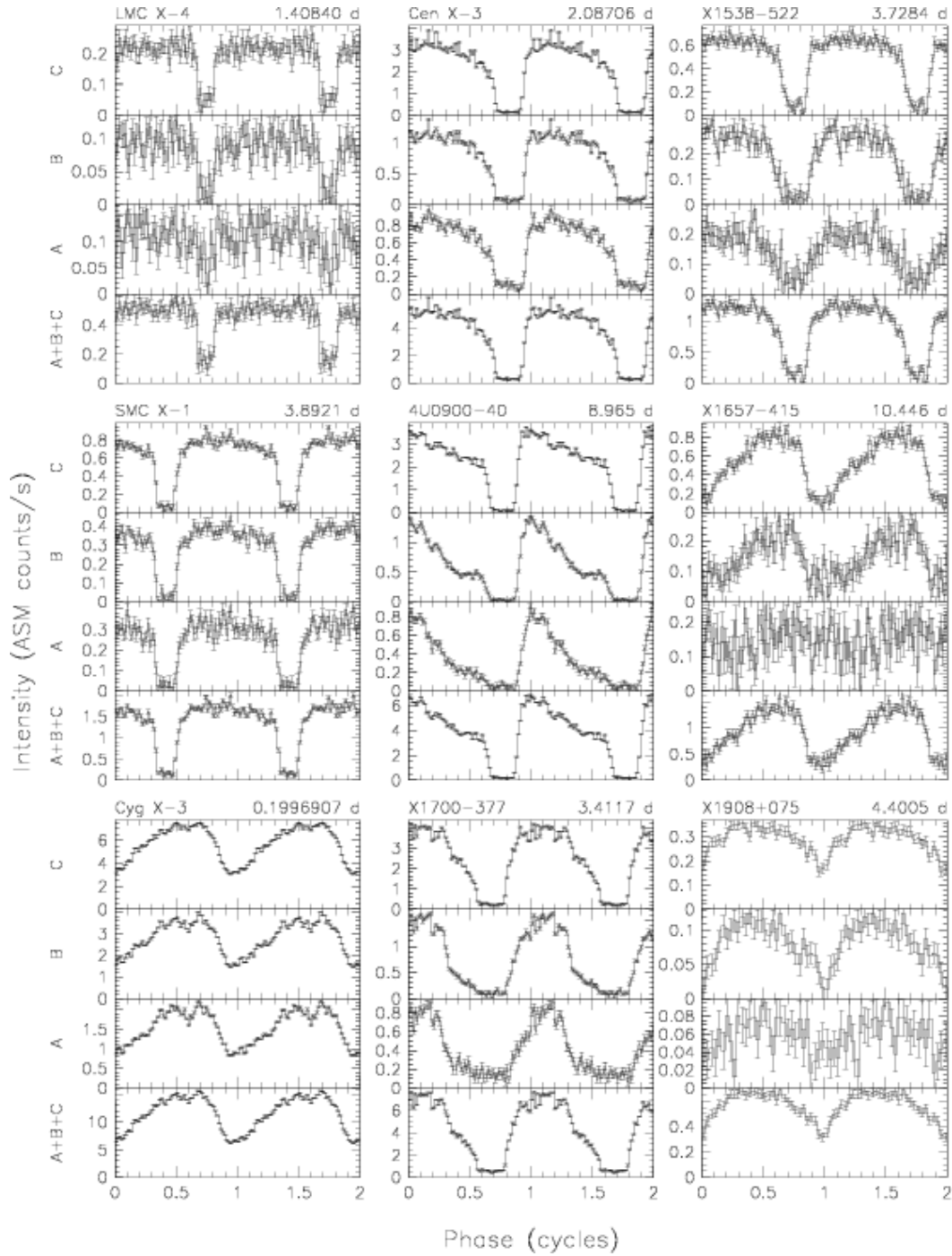}
\caption{Light curves for HMXB supergiant systems folded at the
orbital periods from Table 1.  For each source, the folded light curve
is given for the 1.5--3 keV (A), 3--5 keV (B), 5--12 keV (C), and
1.5--12 keV (A+B+C) bands.}\label{fold_2}
\end{figure}

\begin{figure}
\includegraphics[clip, width=5.5in]{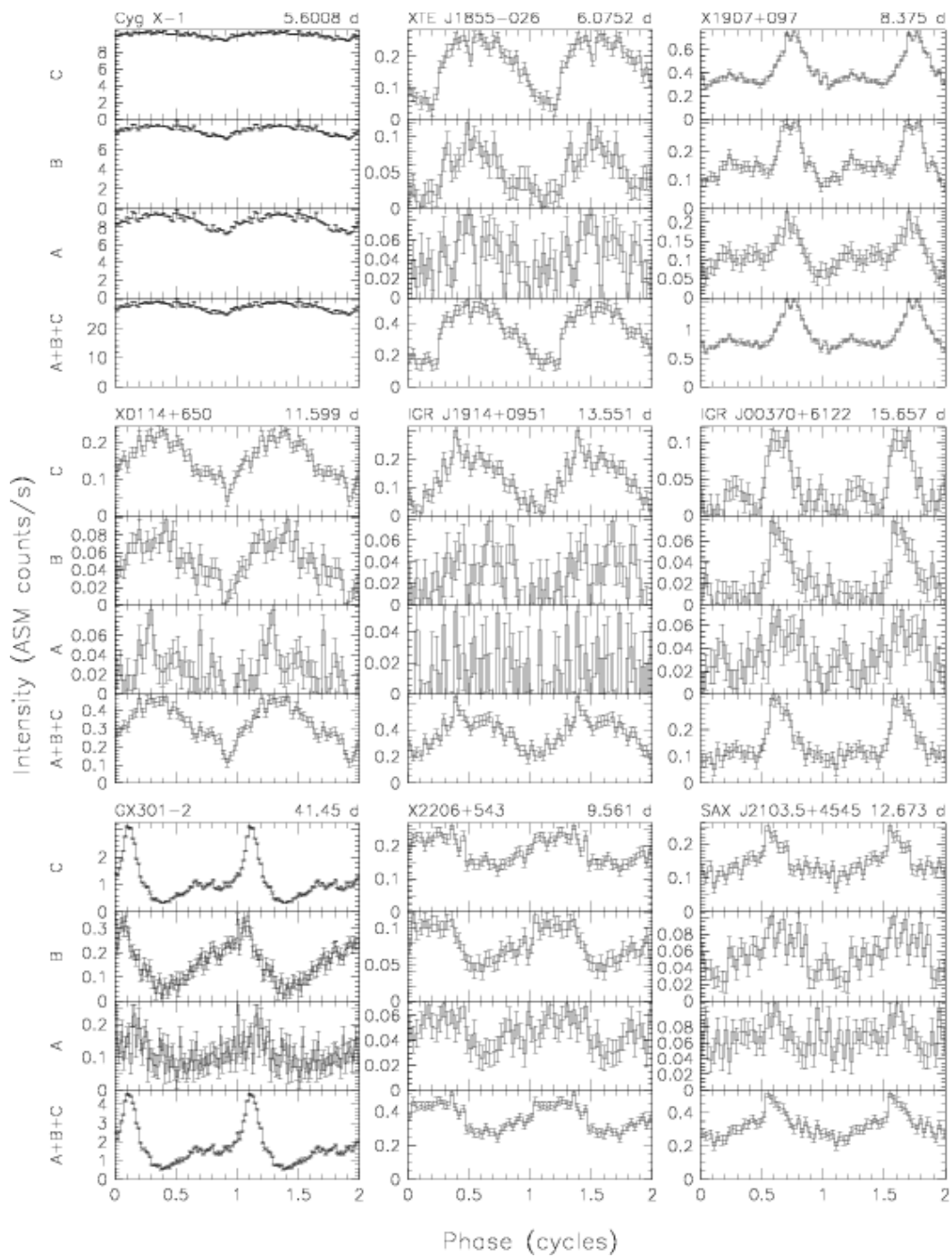}
\caption{Light curves for HMXB supergiant and Be-star systems folded at the
orbital periods from Table 1.  For each source, the folded light curve
is given for the 1.5--3 keV (A), 3--5 keV (B), 5--12 keV (C), and
1.5--12 keV (A+B+C) bands.}\label{fold_3}
\end{figure}

\begin{figure}
\includegraphics[clip, width=5.5in]{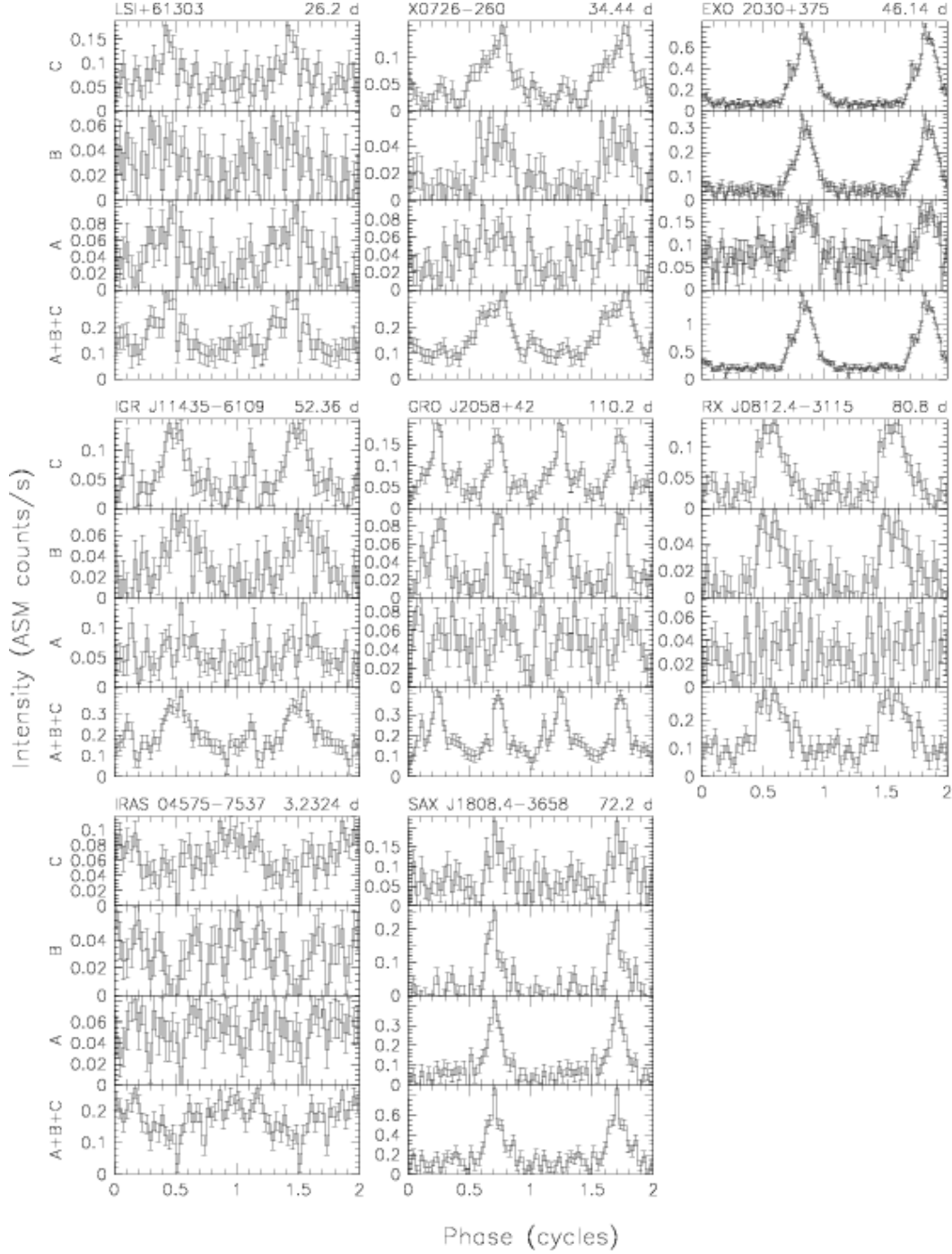}
\caption{Light curves for HMXB Be-star systems and two other systems
with periodicities of marginal statistical significance
(IRAS04575$-$7537 and SAX~J1808.4$-$3658) folded at the orbital
periods from Tables 1 and 3 (see text) except for the light curve of GRO
J2058+42 which is folded using a period that is twice that in Table 1.
For each source, the folded light curve is given for the 1.5--3 keV
(A), 3--5 keV (B), 5--12 keV (C), and 1.5--12 keV (A+B+C)
bands.}\label{fold_4}
\end{figure}

\begin{figure}
\includegraphics[clip, width=5.5in]{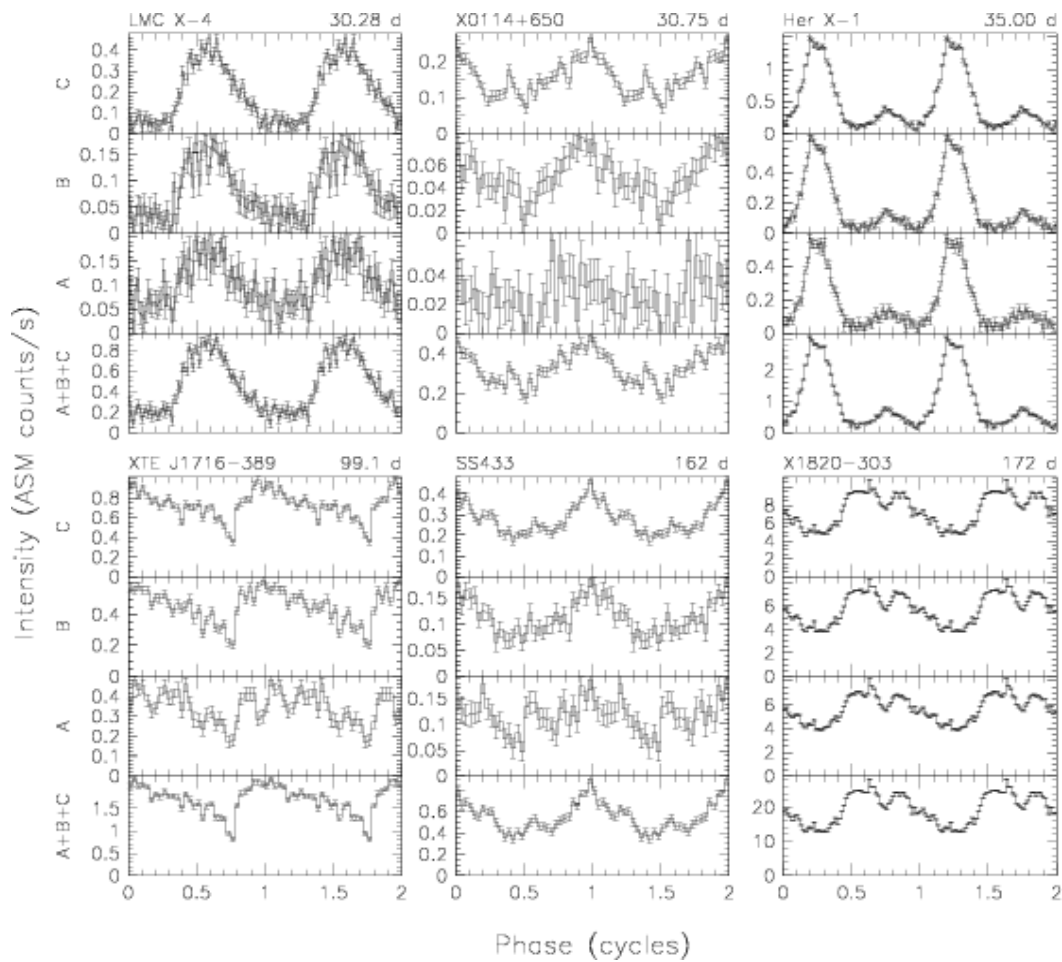}
\caption{Light curves for five superorbital periodicities and one
periodicity of undetermined origin (XTE J1716$-$389) folded at the
orbital periods from Table 2.  For each source, the folded light curve
is given for the 1.5--3 keV (A), 3--5 keV (B), 5--12 keV (C), and
1.5--12 keV (A+B+C) bands.}\label{fold_5}
\end{figure}

\plotone{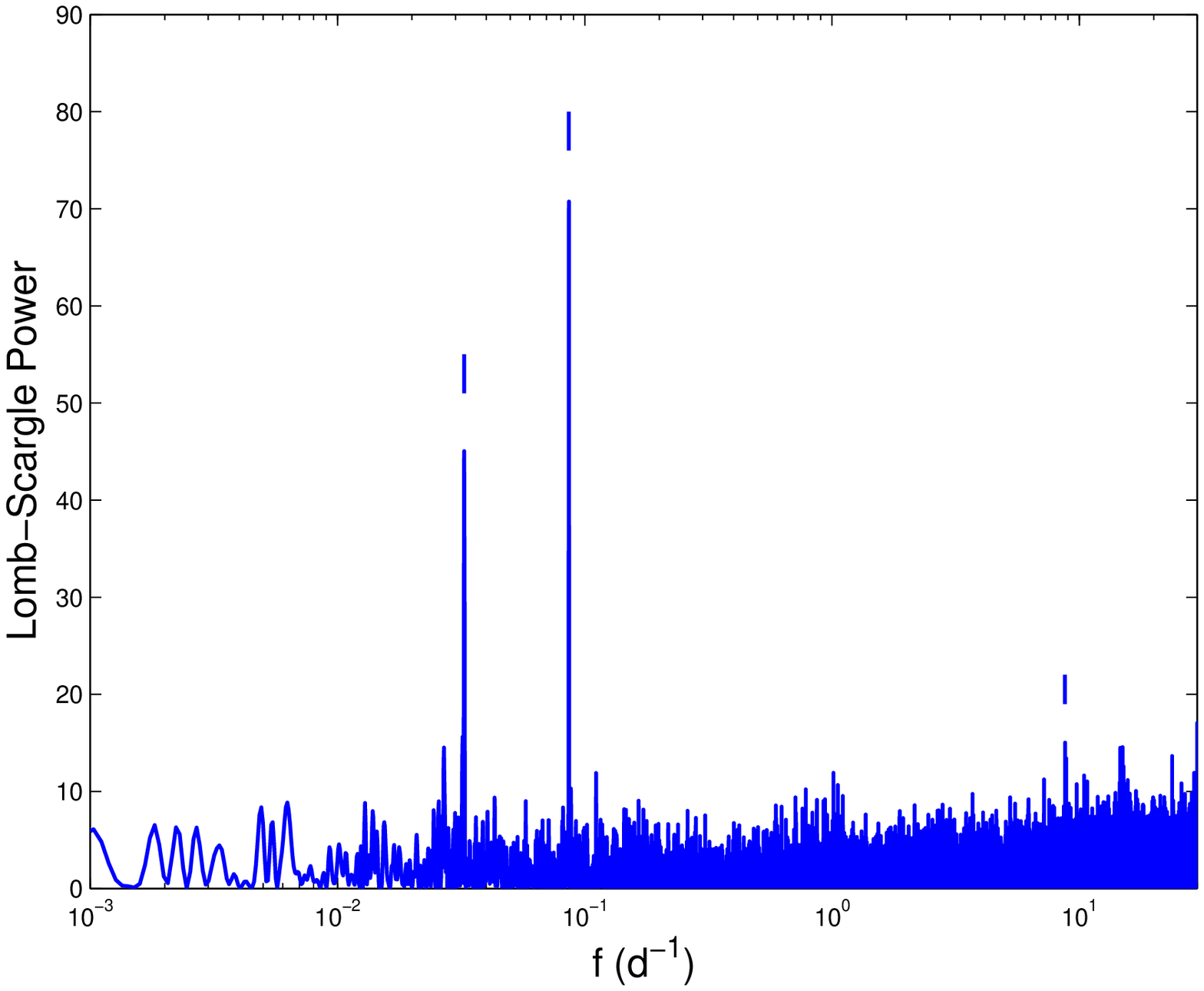}
\begin{figure}
\caption{Lomb-Scargle periodogram from 8.5 years of 90-s time
resolution ASM data in the 1.5--12 keV band on X0114+650. The 30.8 d,
11.6d, and 2.74 h periods are indicated with short vertical
lines.}\label{lomb_x0114}
\end{figure}

\plotone{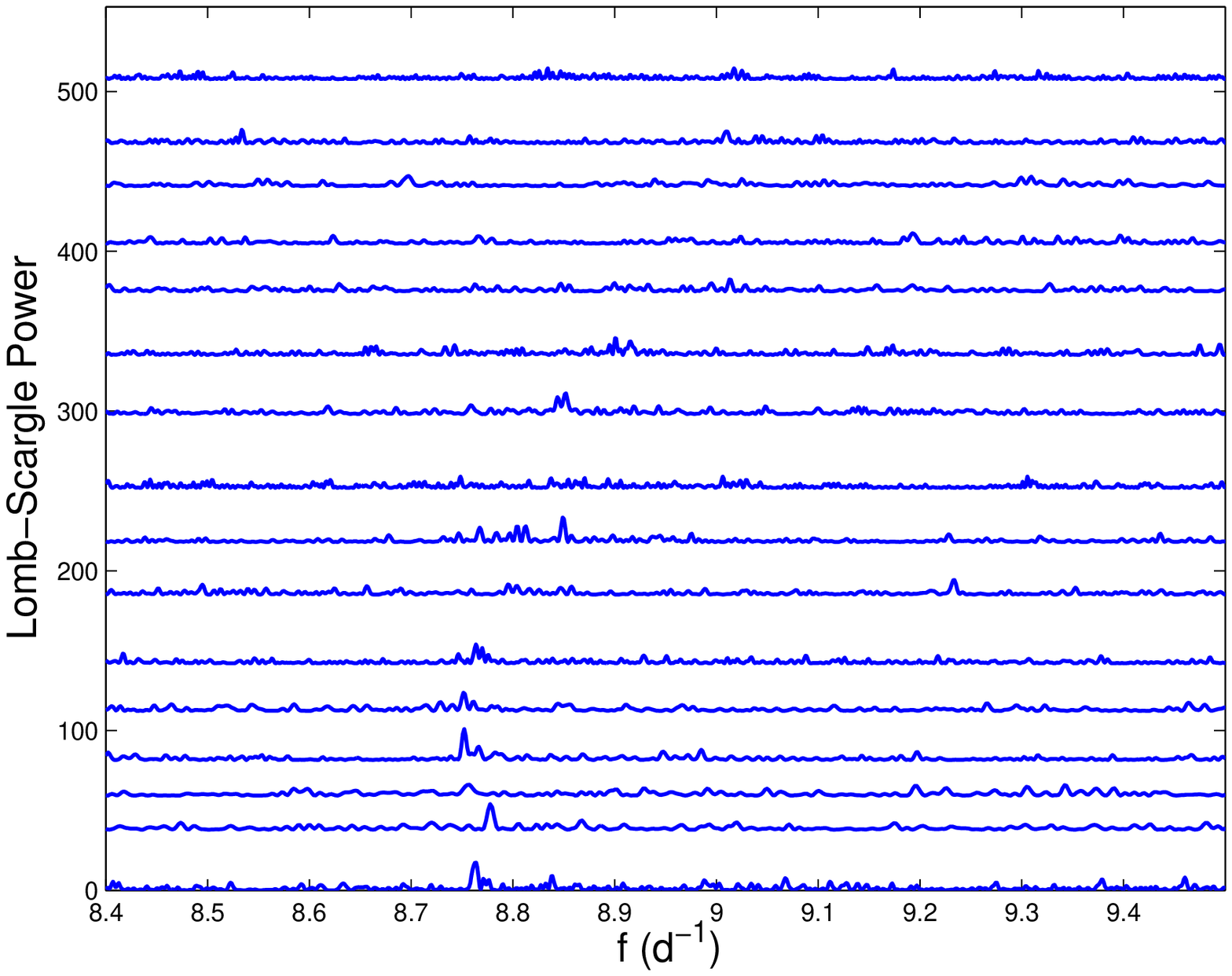}
\begin{figure}
\caption{Lomb-Scargle periodograms (1.5--12 keV band) for roughly 0.5 year
intervals of X0114+650. The periodograms are displaced upward, i.e.,
parallel to the power axis, by an amount proportional to the time of
the beginning of the interval after the initial time of the first
interval. The
period (or frequency) changes of the signal near a period of 2.74 h
(around 8.8 d$^{-1}$) is apparent.}\label{t_f_x0114_1}
\end{figure}

\plotone{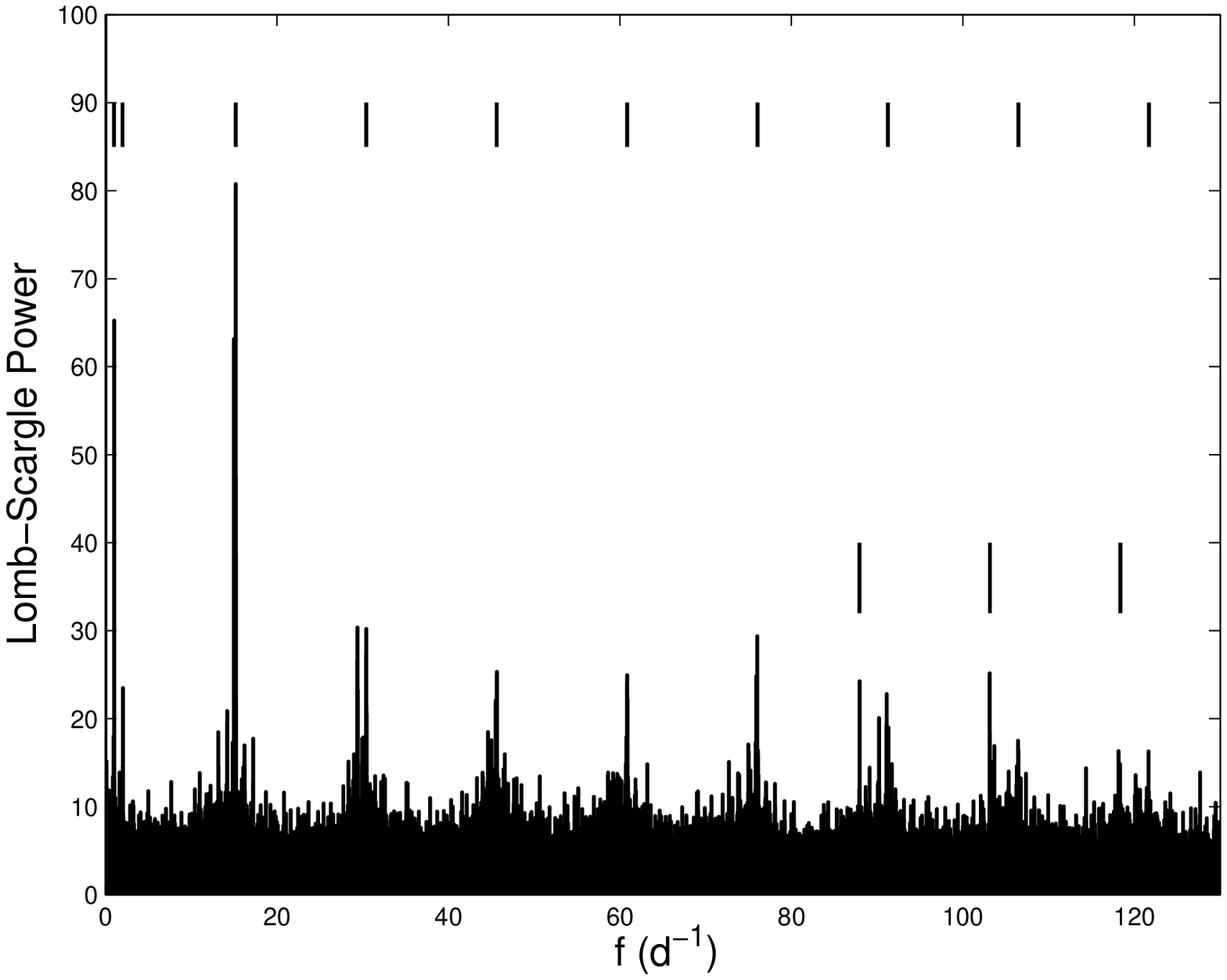}
\begin{figure}
\caption{Lomb-Scargle periodogram of X0352+30 (X~Per) in the 1.5--12 keV band
made from 8.5 years of 90-s time resolution data.  The 837 s period
($\sim 103$ d$^{-1}$) and its beats with the $\sim 15$ d$^{-1}$
orbital frequency of the spacecraft are indicated with three vertical
lines.  Also indicated with shorter vertical lines are the 15 d$^{-1}$
spacecraft orbital frequency and 7 of its harmonics, as well as the 1
d$^{-1}$ rotation frequency of the Earth and its second
harmonic.}\label{lomb_xper}
\end{figure}

\plotone{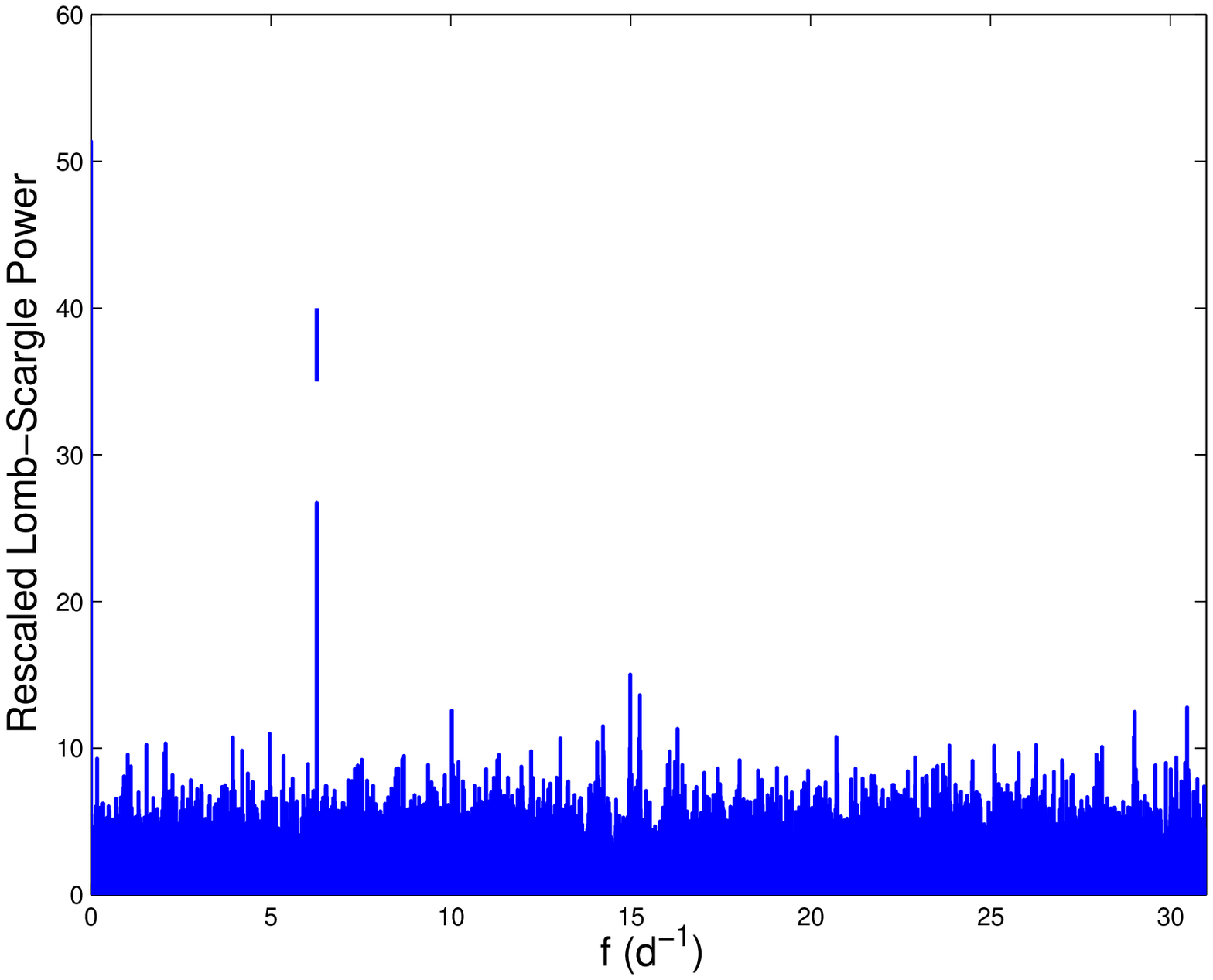}
\begin{figure}
\caption{Rescaled Lomb-Scargle periodogram for EXO0748$-$676 in the
3--5 keV band made from 8.5 years of 90-s time resolution data.  The
3.8 h period is indicated with a short vertical
line.}\label{lomb_exo0748}
\end{figure}

\plotone{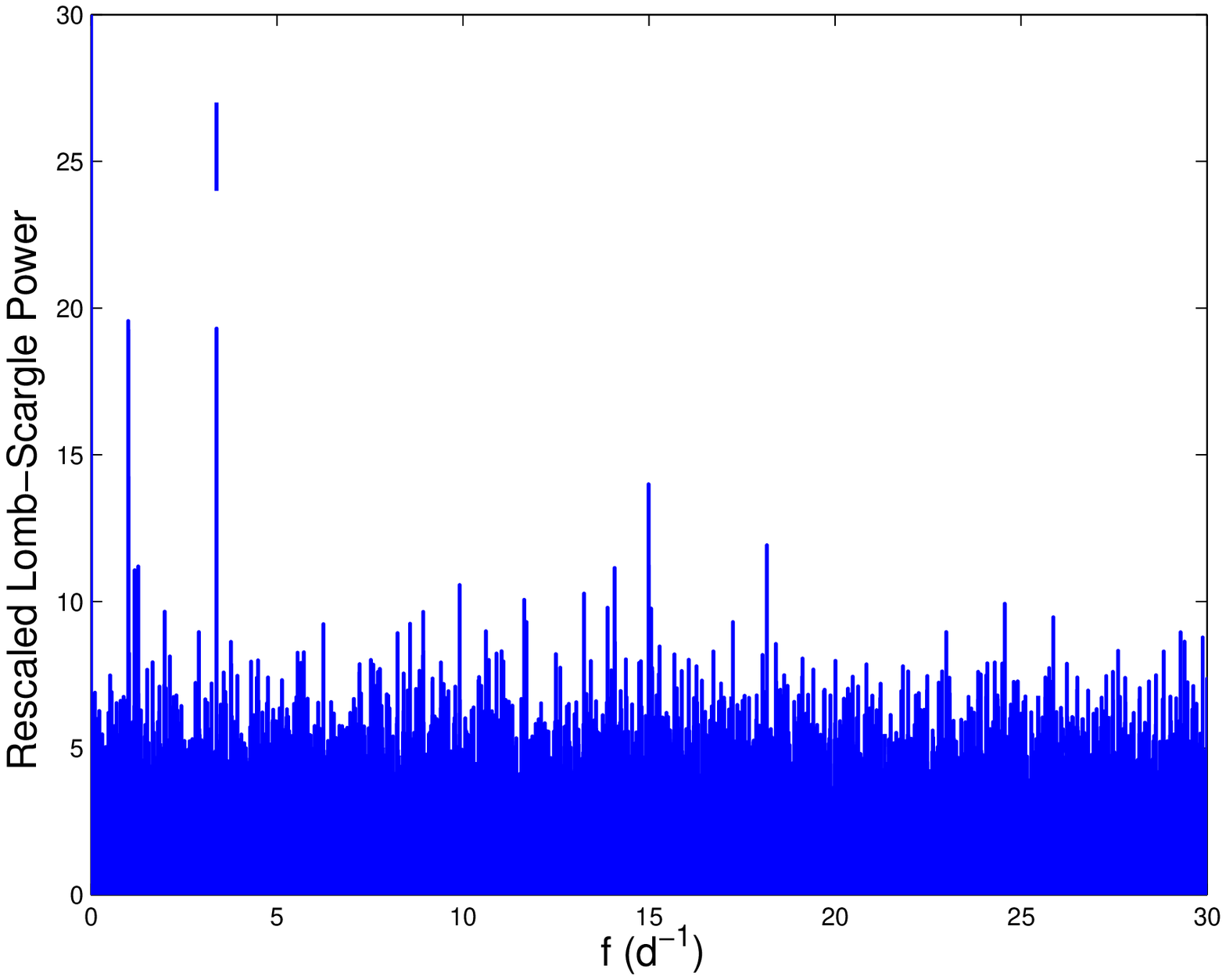}
\begin{figure}
\caption{Rescaled Lomb-Scargle periodogram for X1658$-$298 in the 1.5--12 keV
band made from 6.5 years of 90-s time resolution data.  The 7.11 h
period is indicated with a vertical line.  The other two high powers
are artifacts at 1 d$^{-1}$ and 15 d$^{-1}$. }\label{lomb_x1658}
\end{figure}

\plotone{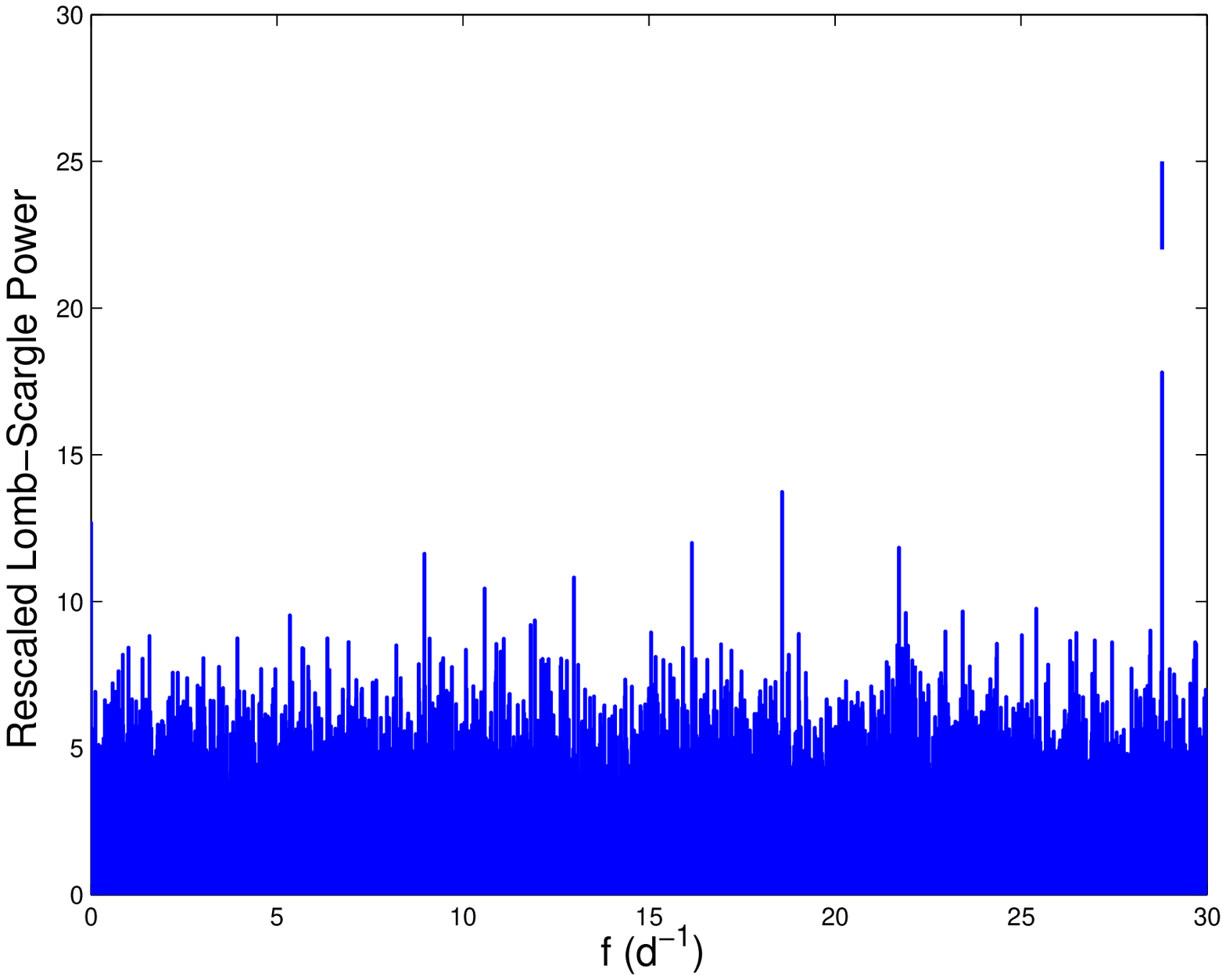}
\begin{figure}
\caption{Rescaled Lomb-Scargle periodogram for X1916$-$053 in the 1.5--12 keV
band made from 6.5 years of 90-s time resolution data.  The 3000 s
period, indicated with a short vertical line, is the highest peak in
the periodogram. }\label{lomb_x1916}
\end{figure}

\plotone{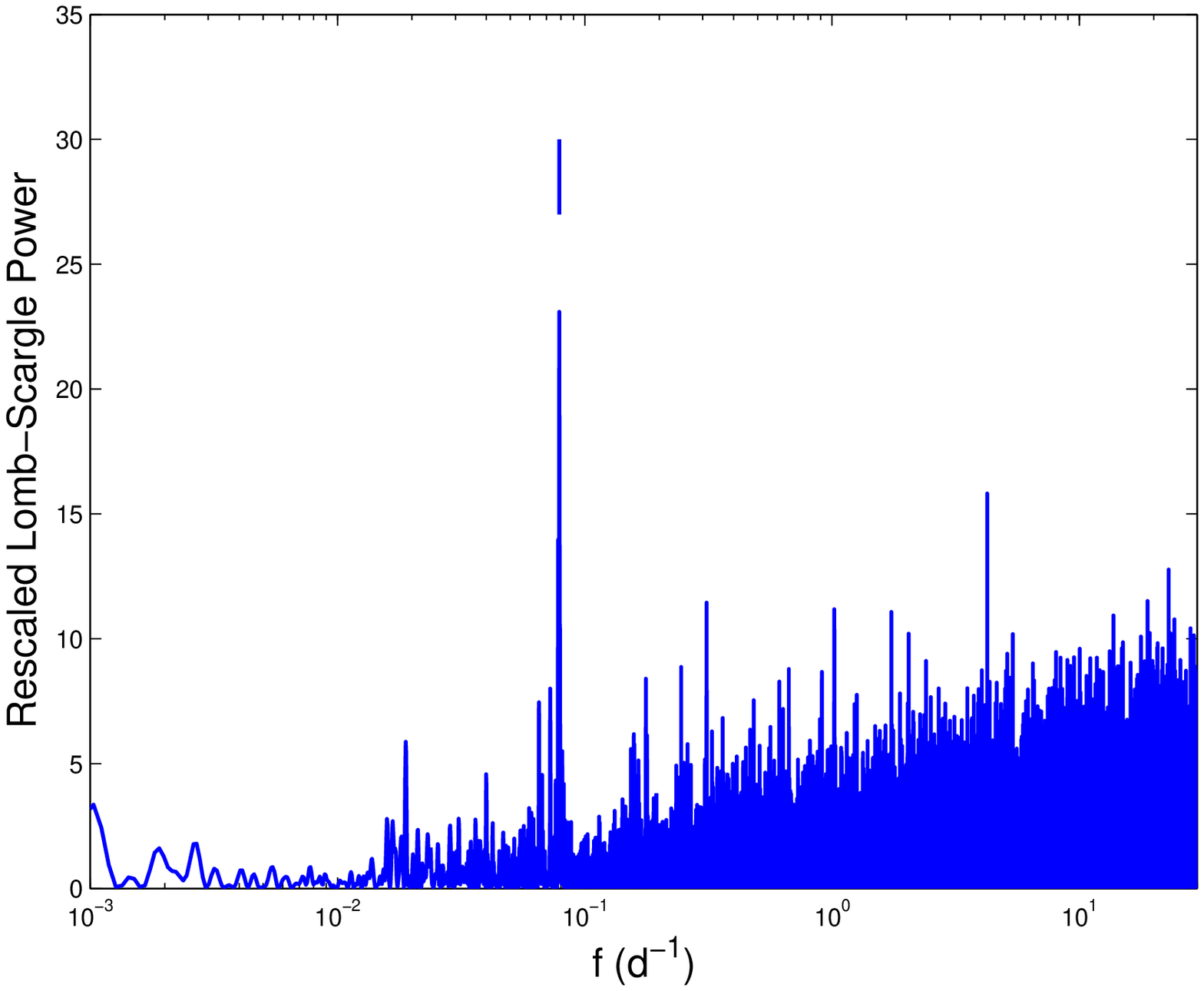}
\begin{figure}
\caption{Rescaled Lomb-Scargle periodogram for SAX~J2103.5+4545 in the 1.5--12
keV band made from 8.5 years of 90-s time resolution data. The 12 d
period is indicated with a vertical line. The second largest peak is
at a 2.86 h period with FAP $\sim0.1$. }\label{lomb_saxj2103}
\end{figure}

\plotone{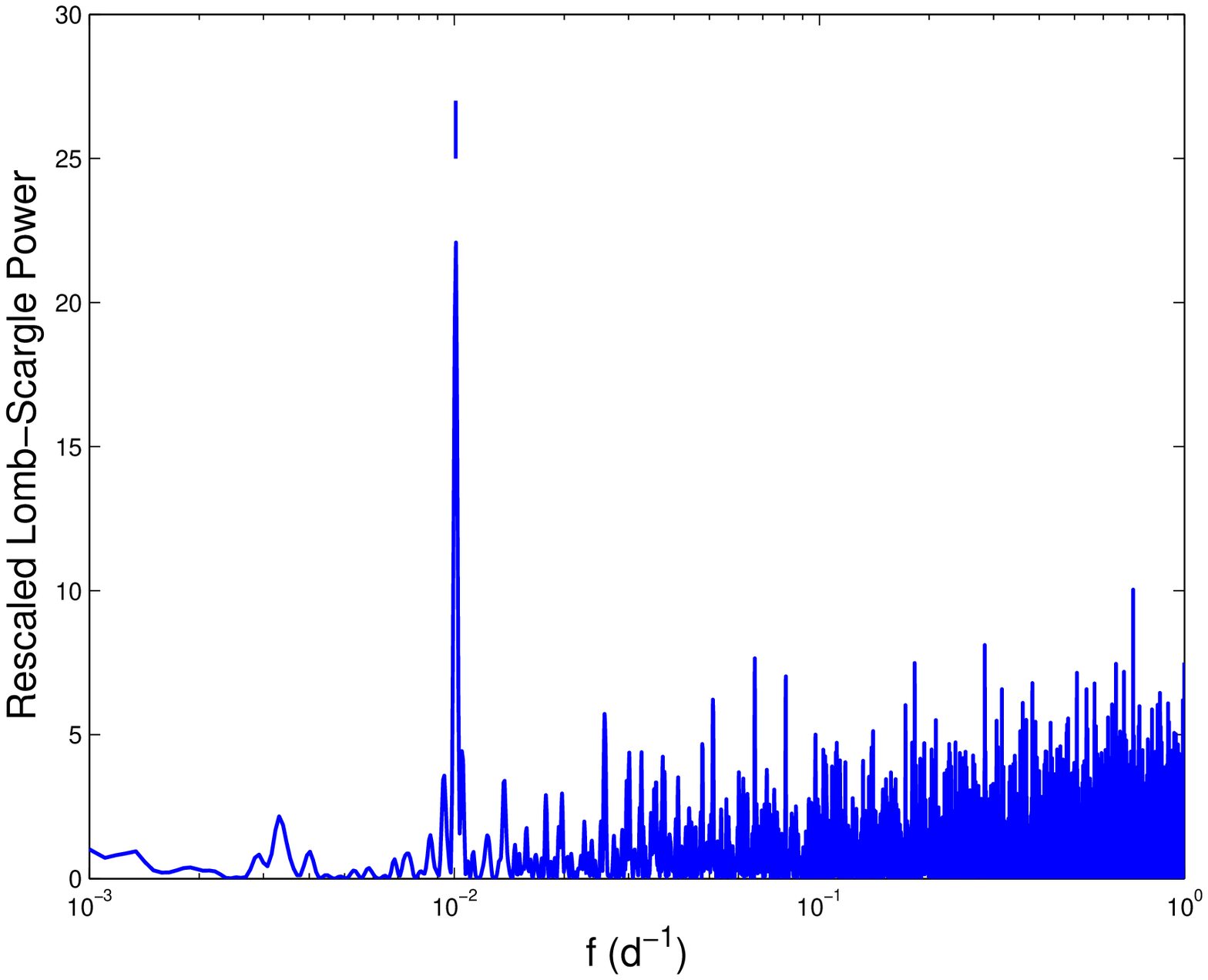}
\begin{figure}
\caption{Rescaled Lomb-Scargle periodogram for XTE~J1716$-$389 in the 1.5--12 keV
band made from 8.5 years of 1-d time averages. The $\sim$98 d peak
is indicated with a vertical line.  }\label{lomb_x1716}
\end{figure}

\plotone{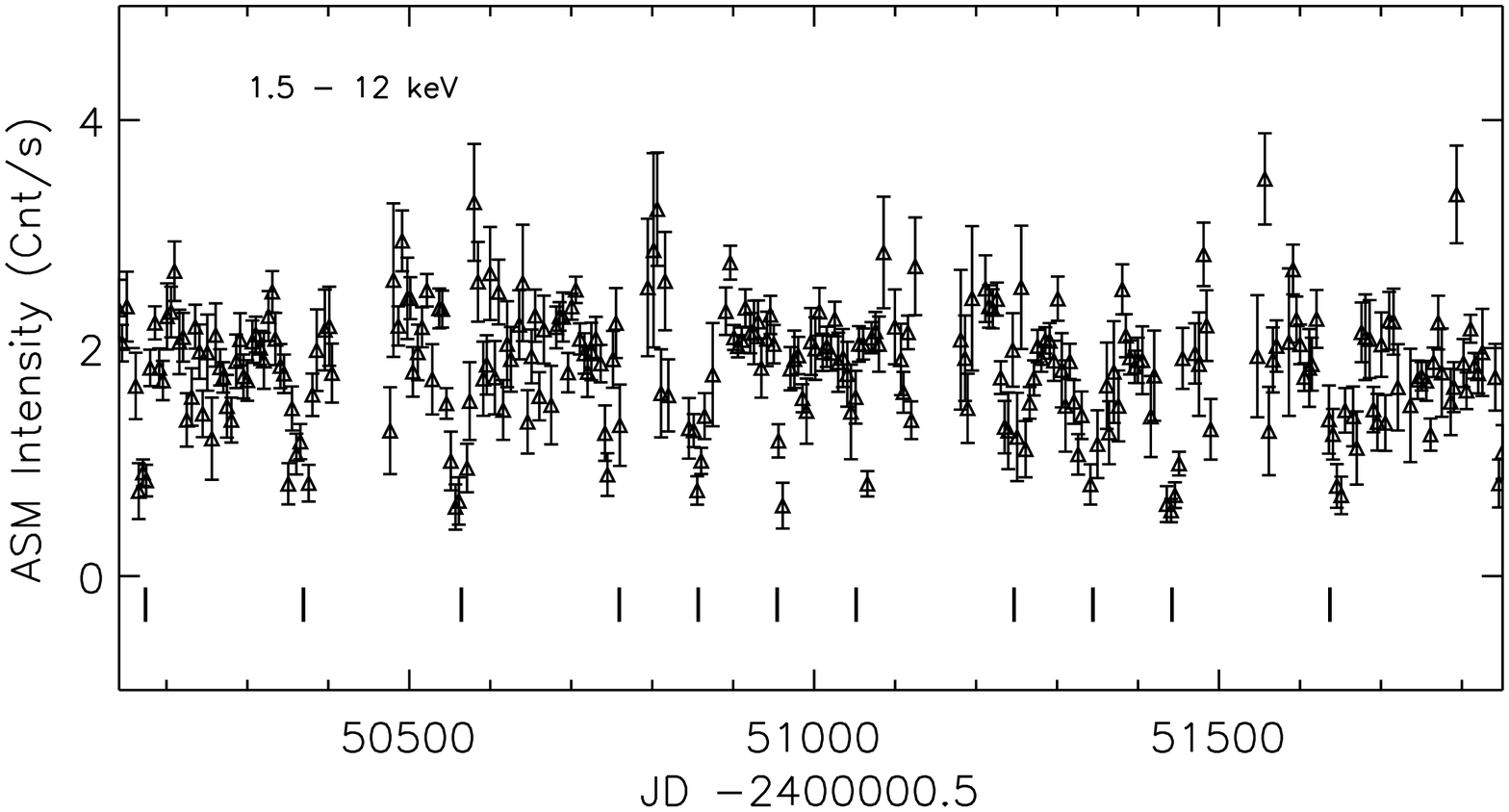}
\begin{figure}
\caption{A section of the 1.5--12 keV ASM light curve of XTE~J1716$-$389
in 5-d time bins.  The visible dips are labeled with short vertical
lines.}\label{ltc_x1716_asm}
\end{figure}

\plotone{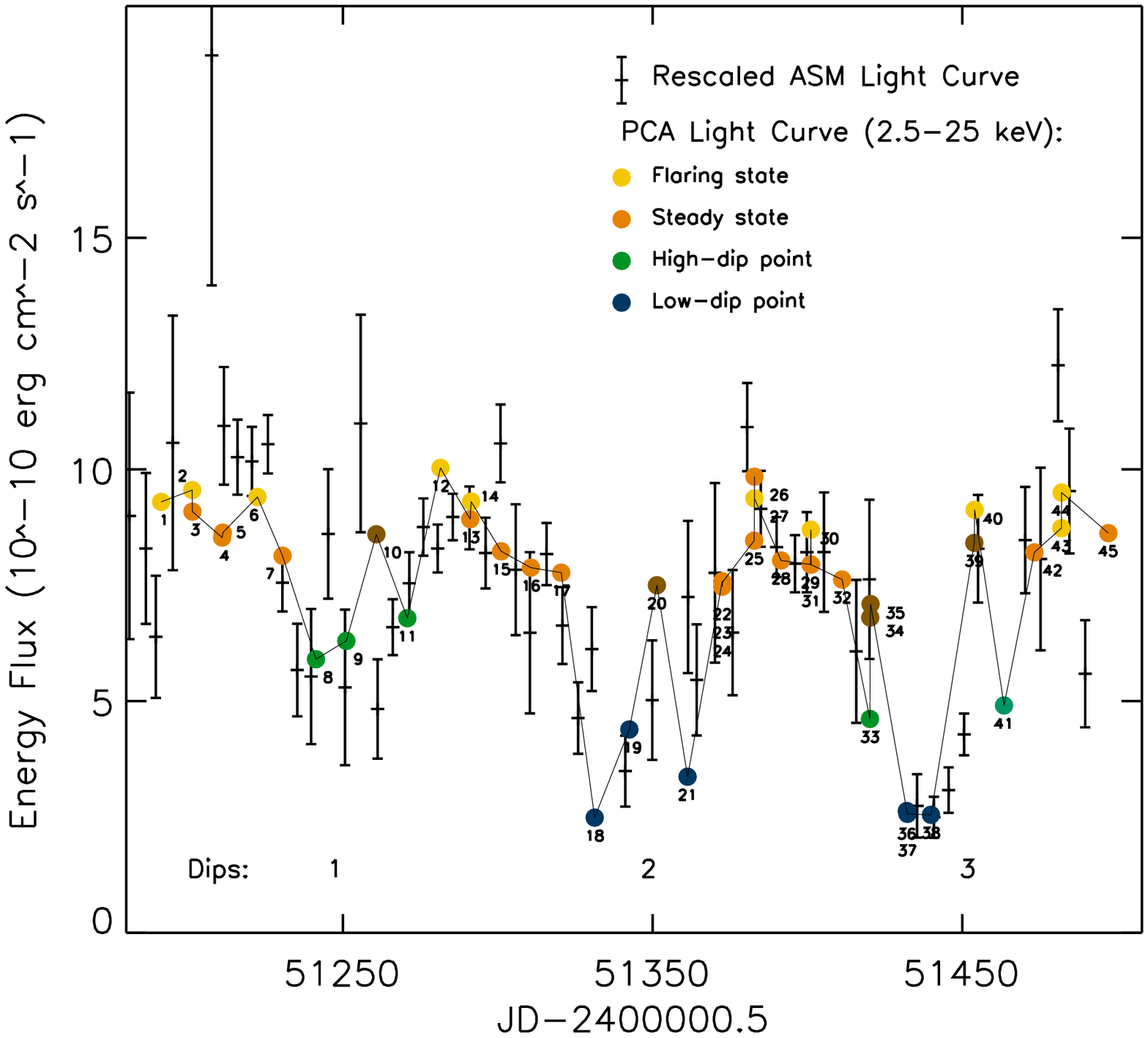}
\begin{figure}
\caption{The 2.5--25 keV PCA energy flux of XTE~J1716$-$389 as a
function of time and a rescaled ASM light curve in 5-d time bins.
The observation sequences are labeled.  Darker shades of the filled circles
indicate increasing amount of absorption column density required for
spectral fitting in the PCA data. Three broad dips are apparent
in both light curves. The PCA data further confirm the detection of
the periodic dips in this source.}\label{ltc_x1716_asm_pca}
\end{figure}

\plotone{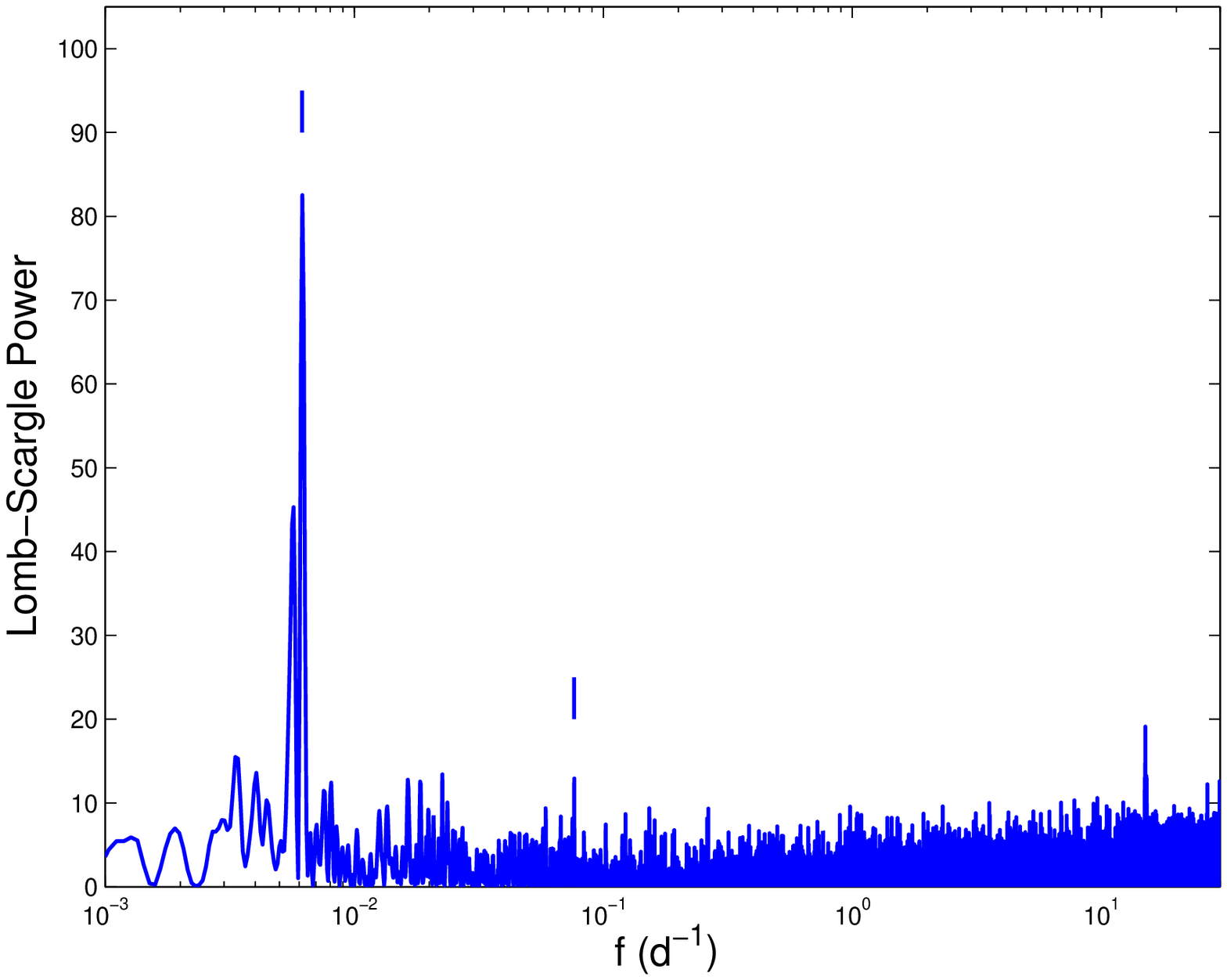}
\begin{figure}
\caption{Lomb-Scargle periodogram for SS~433 in the 5--12 keV band
made from 8.5 years of 90-s time resolution data.  The 162 d
precession period and the 13.1 d orbital period are indicated with
vertical lines.  }\label{lomb_ss433}
\end{figure}

\plotone{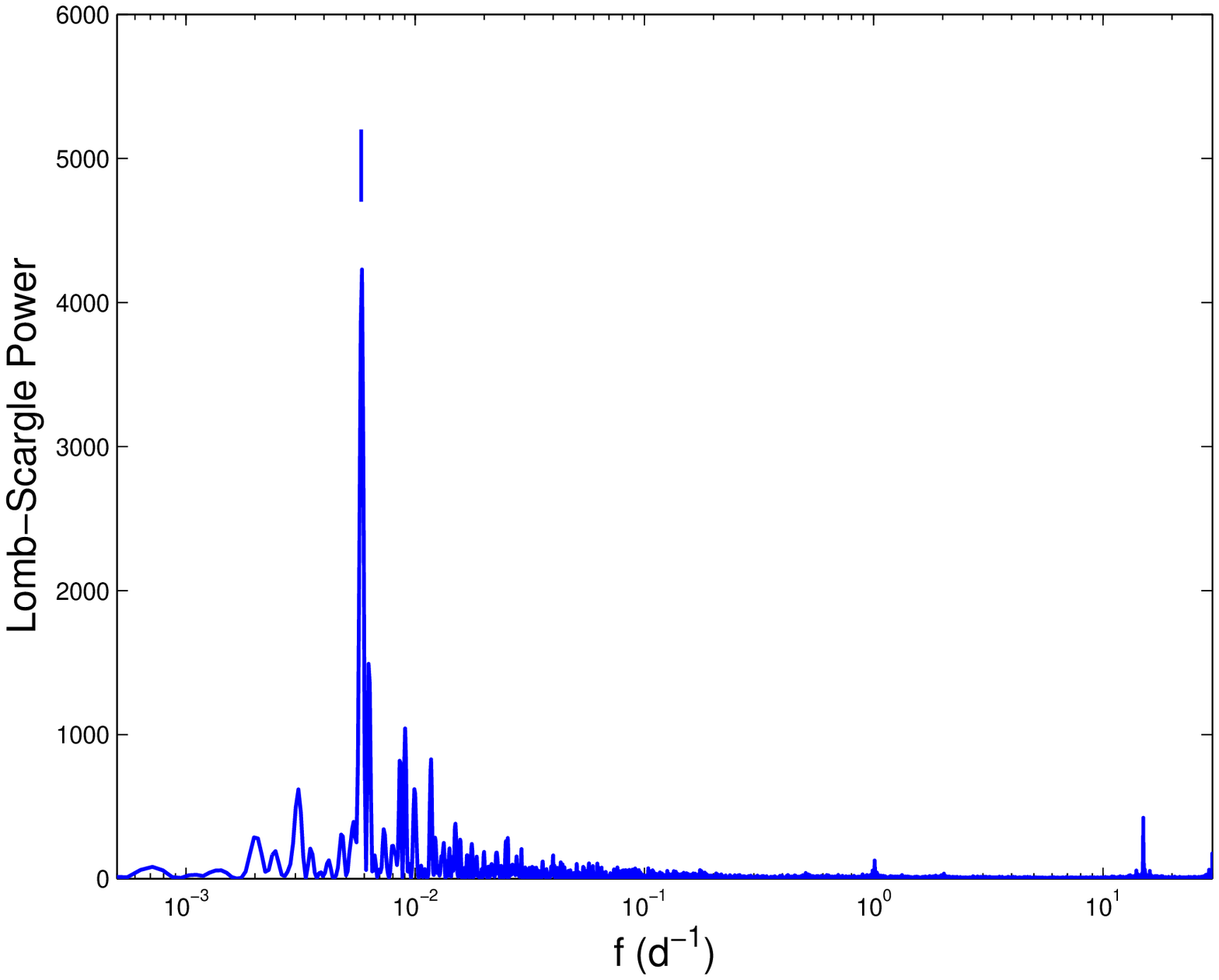}
\begin{figure}
\caption{Lomb-Scargle periodogram for X1820$-$303 in the 1.5--12 keV
band made from 8.5 years of 90-s time resolution data.  The 172 d
period is indicated.}\label{lomb_x1820}
\end{figure}

\plotone{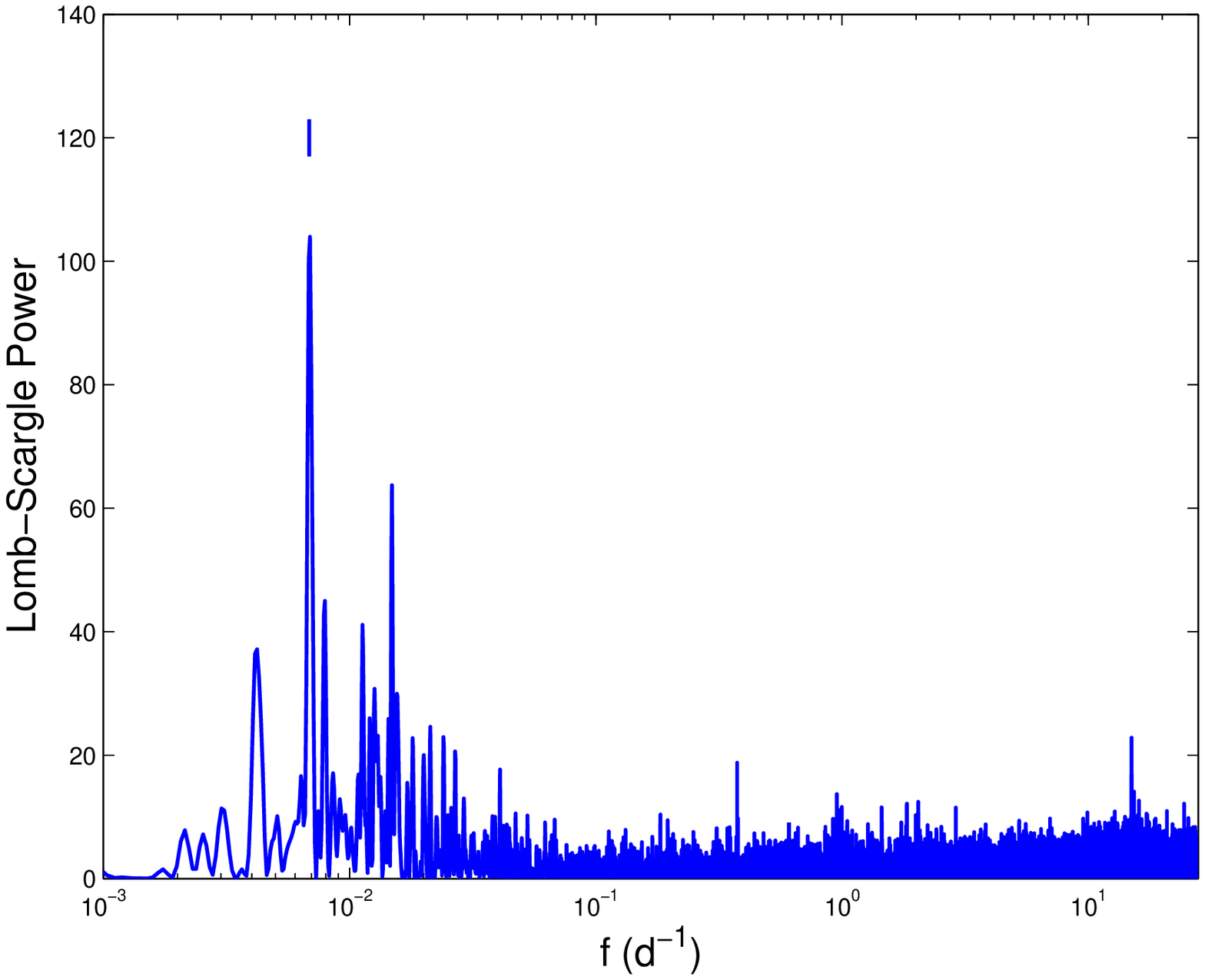}
\begin{figure}
\caption{Lomb-Scargle periodogram for GRS1747$-$312 in the 1.5--12 keV
band in  90-s time bins.  The
quasi-period of 146 d is indicated. }\label{lomb_grs1747}
\end{figure}

\plotone{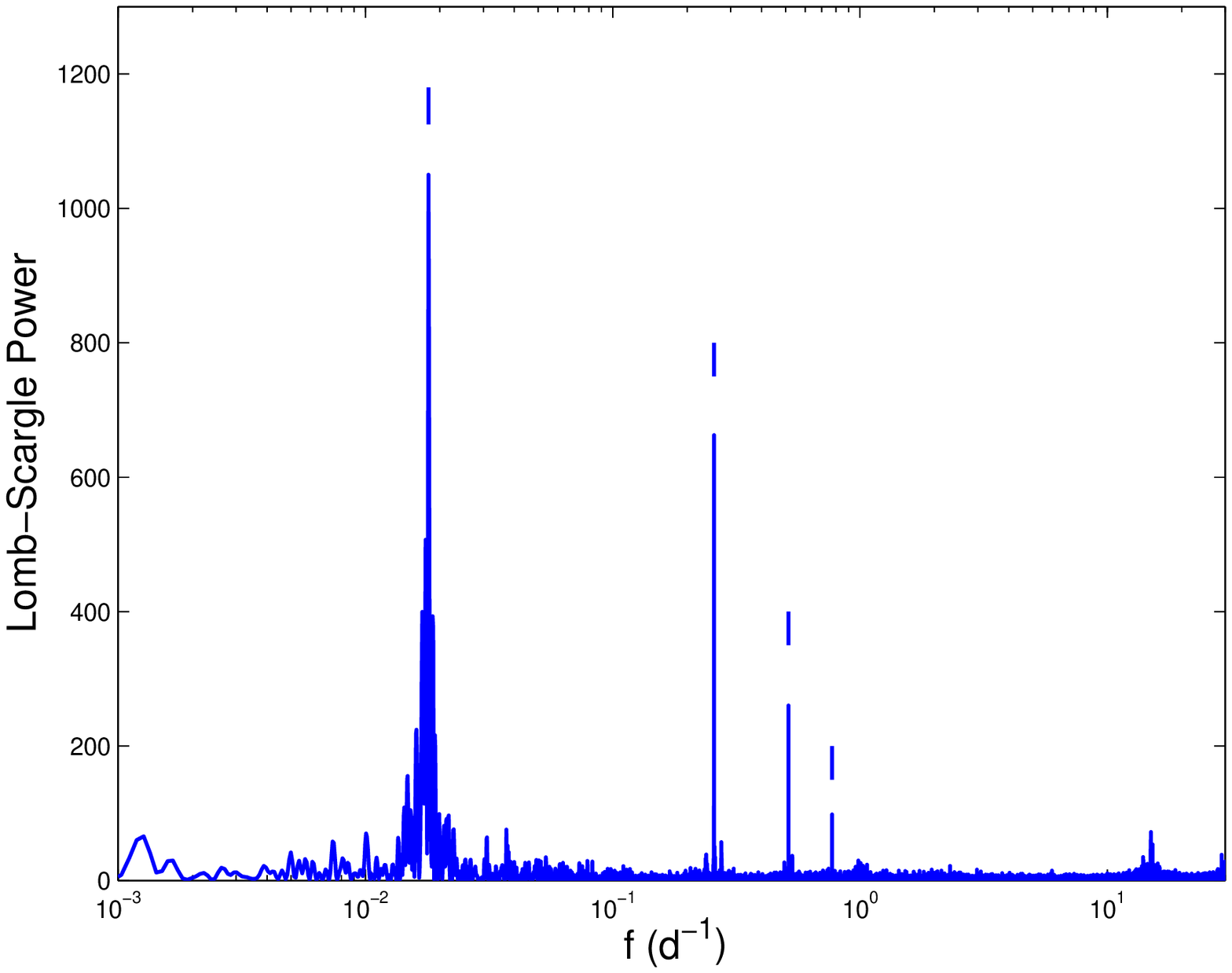}
\begin{figure}
\caption{Lomb-Scargle periodogram for SMC~X-1 in the 1.5--12 keV band
made from 8.5 years of 90-s time resolution data. The detection of
the $\sim 60$ d period and the 3.9 d orbital periods and their
harmonics are clear.  There are multiple peaks and broad features on
time scales of 50--70 d.}\label{lomb_smcx1}
\end{figure}

\plotone{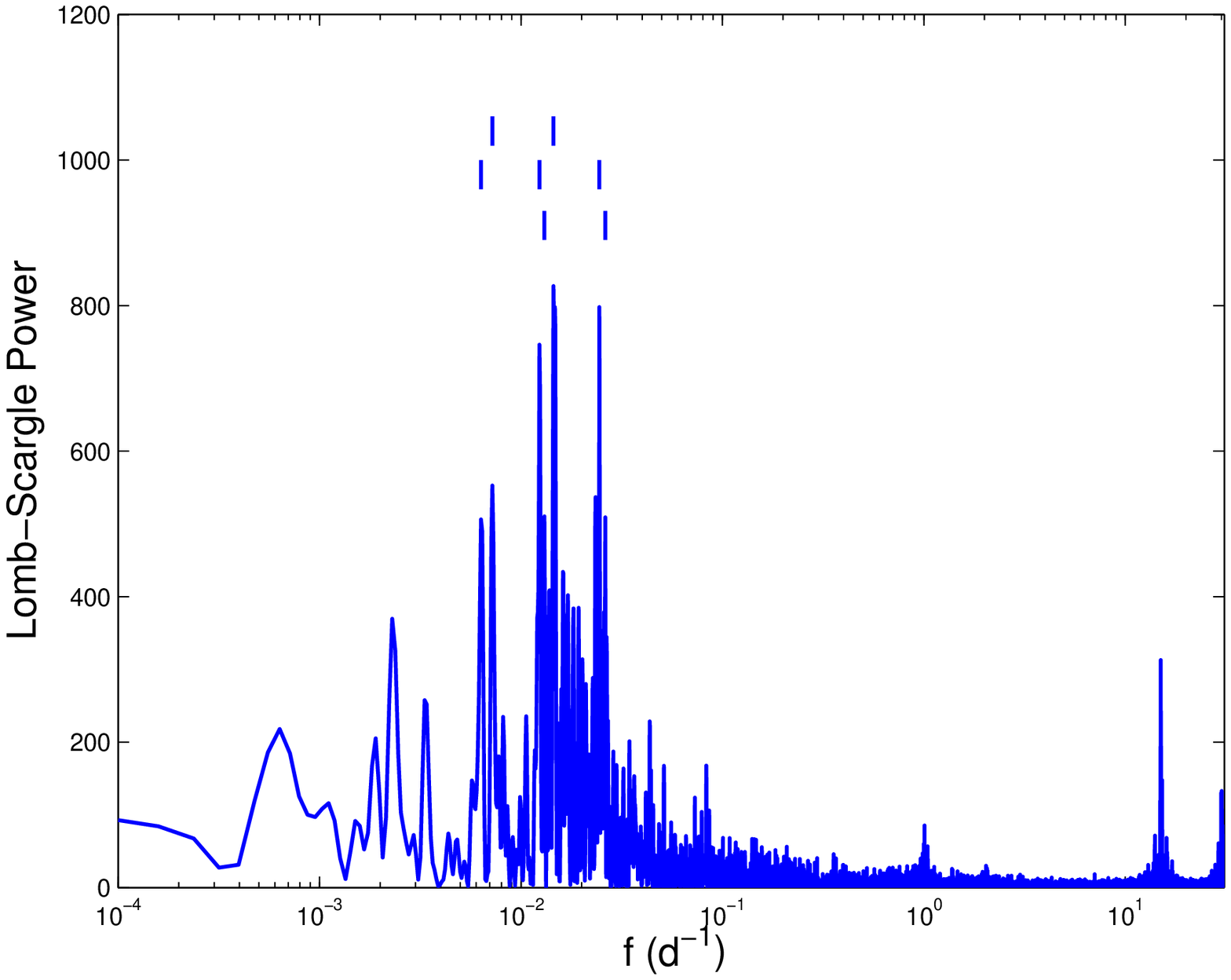}
\begin{figure}
\caption{Lomb-Scargle periodogram for Cyg~X-2 in the 1.5--12 keV band
made from 8.5 years of 90-s time resolution data. There are multiple
peaks at frequencies $<$ 0.03 cycles per day. Three possible sets of
periodicities at 76.5 d, 138.7 d, and 157 d with harmonics are
indicated with short vertical lines. The peaks at 1 d$^{-1}$ and 15
d$^{-1}$ are artifacts (see text).}\label{lomb_cygx2}
\end{figure}

\plotone{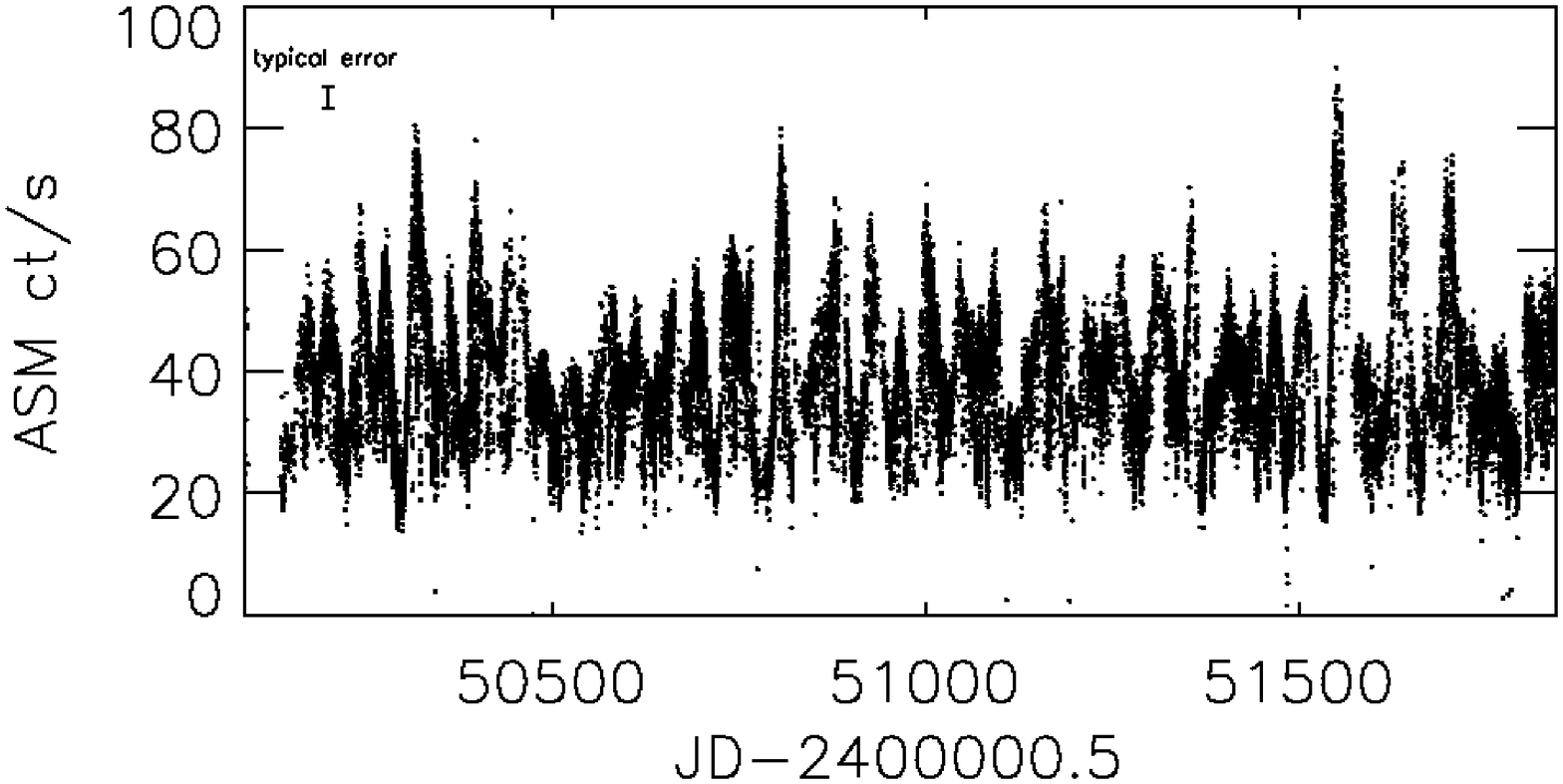}
\begin{figure}
\caption{The ASM 1.5--12 keV band light curve of
Cyg~X-2. Quasi-periodic intensity variations on time scales between
60--90 days are clear.}\label{ltc_cygx2}
\end{figure}

\plotone{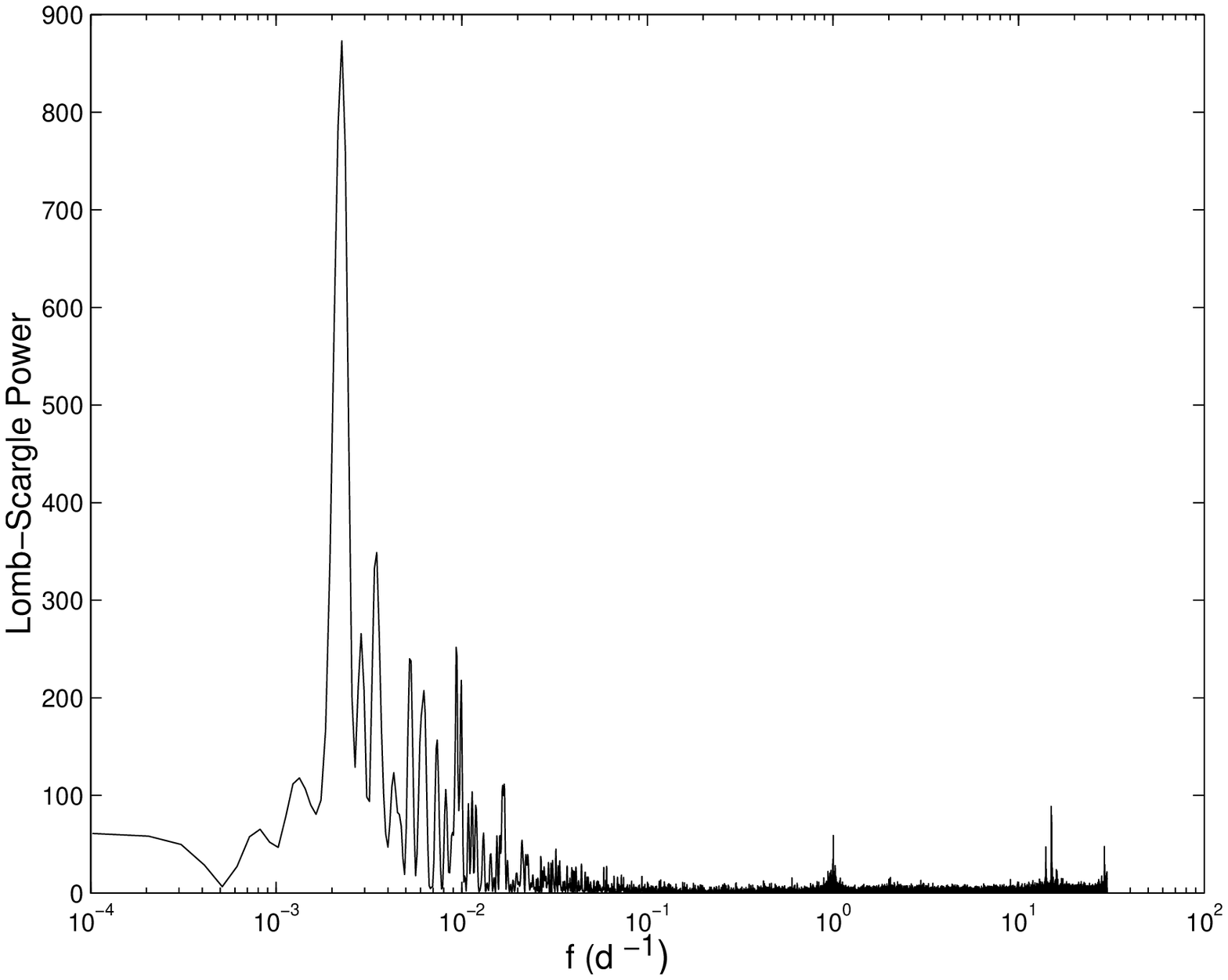}
\begin{figure}
\caption{Lomb-Scargle periodogram for LMC~X-3
 in the {1.5--12} keV band made from 8.5 years of 90-s time resolution
data. Multiple peaks on time scales of 100 to 500 d are
apparent.}\label{lomb_lmcx3}
\end{figure}

\plotone{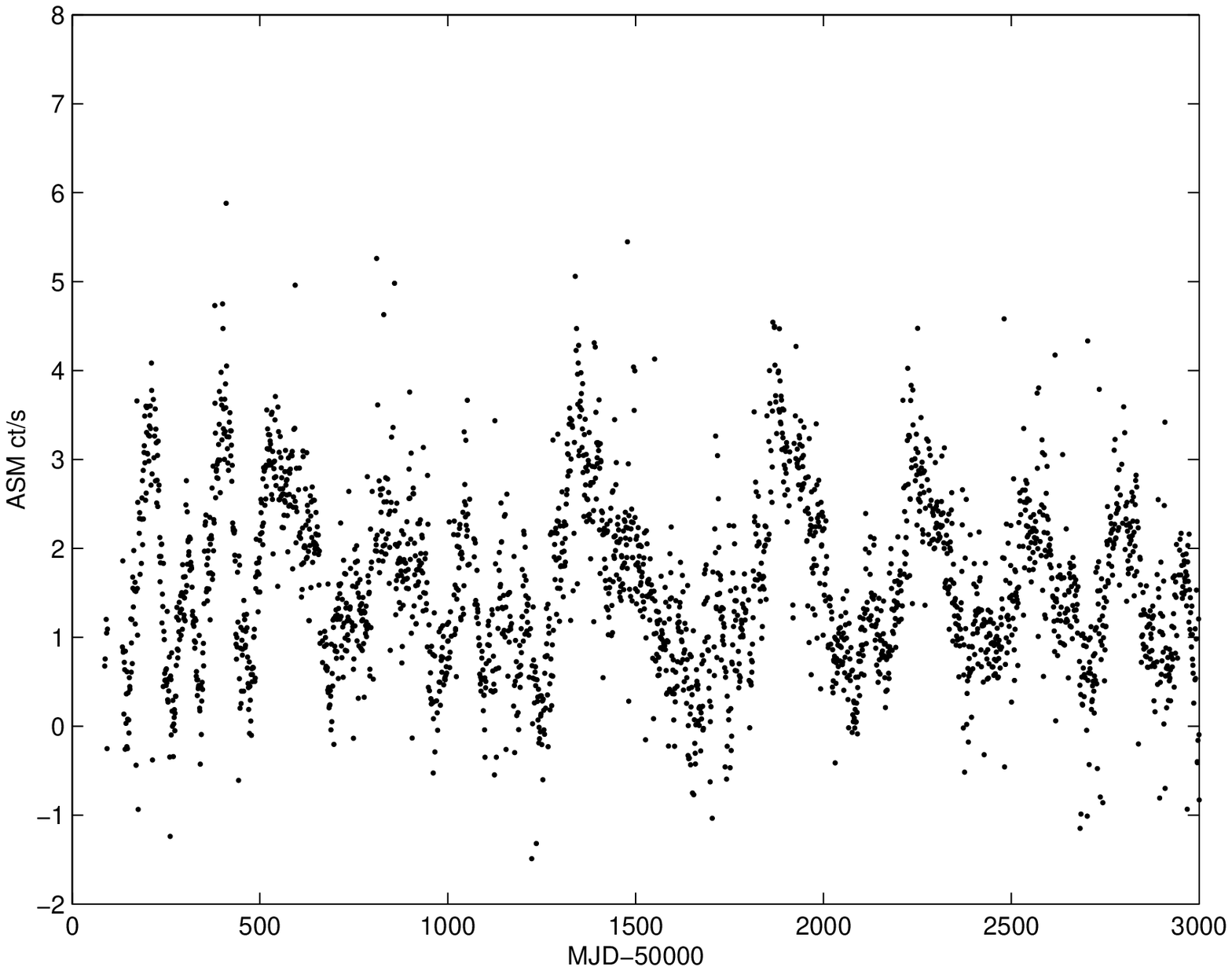}
\begin{figure}
\caption{The ASM 1.5--12 keV band light curve of LMC~X-3. Variability
on time scales of 100--500 d is apparent.}\label{ltc_lmcx3}
\end{figure}

\plotone{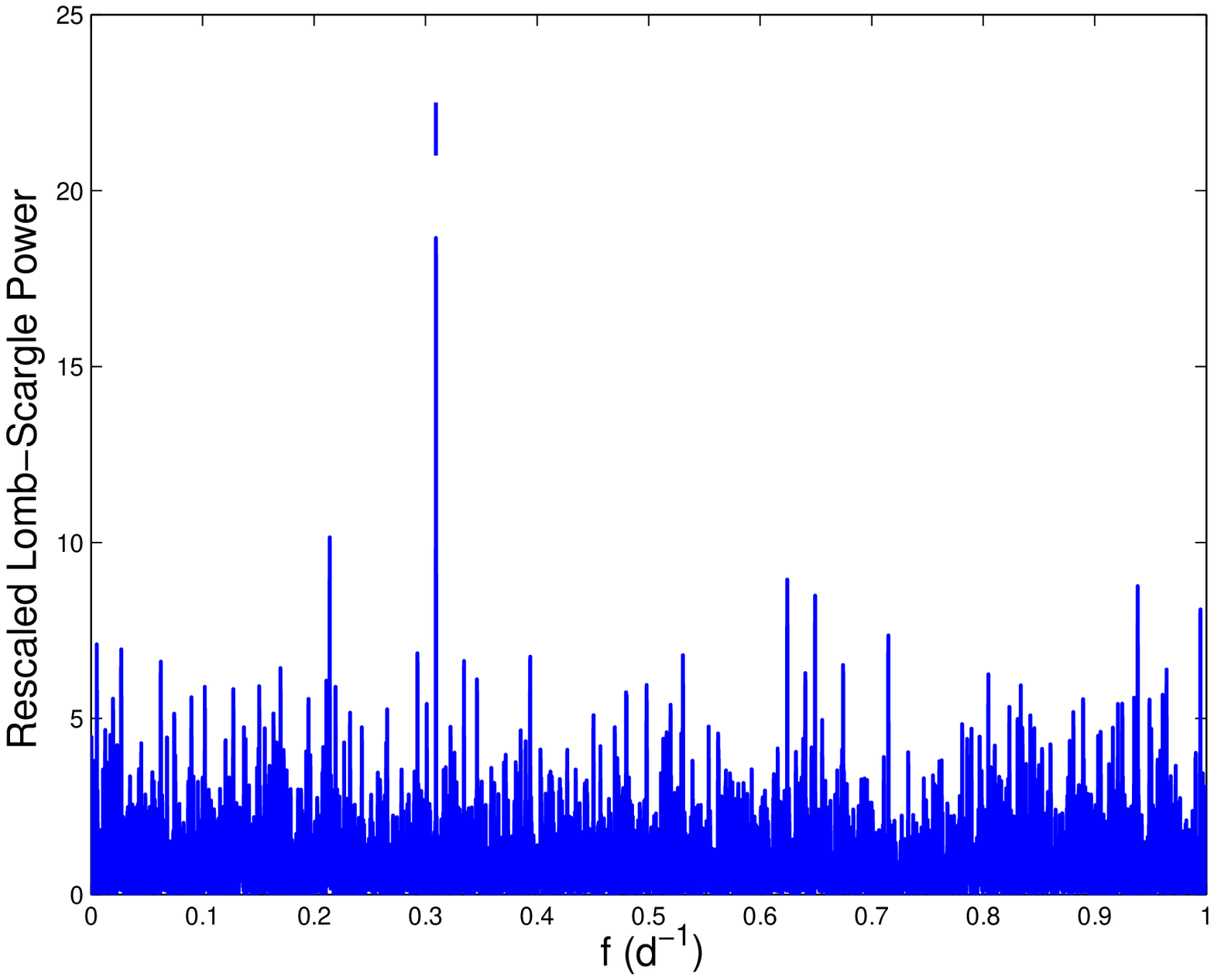}
\begin{figure}
\caption{Rescaled Lomb-Scargle periodogram for IRAS04575$-$7537 in the 1.5--12
keV band made from 8.5 years of 1-d time averages.  The 3.23 d period
is indicated. }\label{lomb_iras04575}
\end{figure}

\plotone{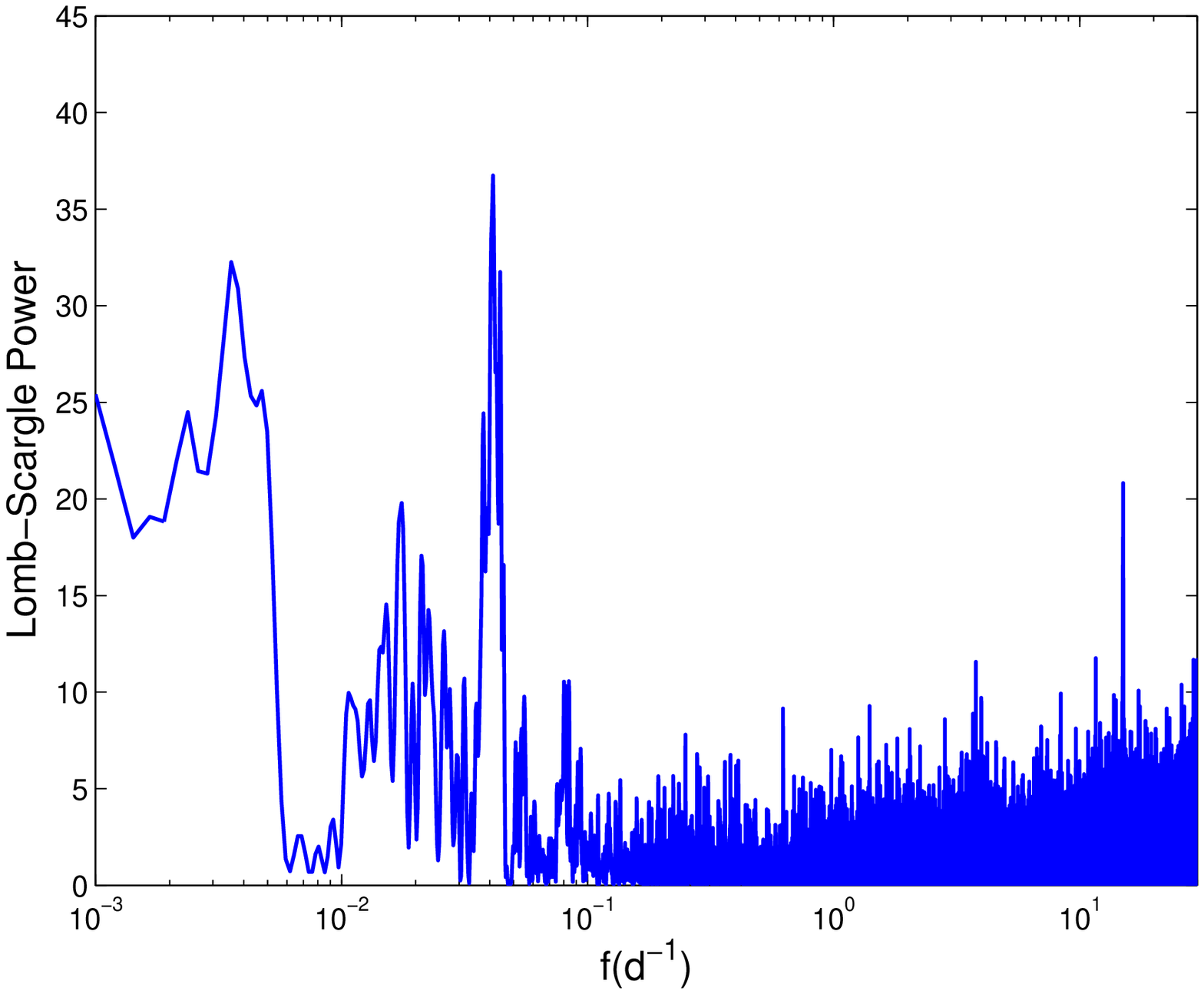}
\begin{figure}
\caption{Lomb-Scargle periodogram for X0115$-$634 in the 3--5 keV
band for the time interval of MJD~50087--51141 (before the first outburst). The
24 d known orbital period is visible.}\label{x0115}
\end{figure}

\plotone{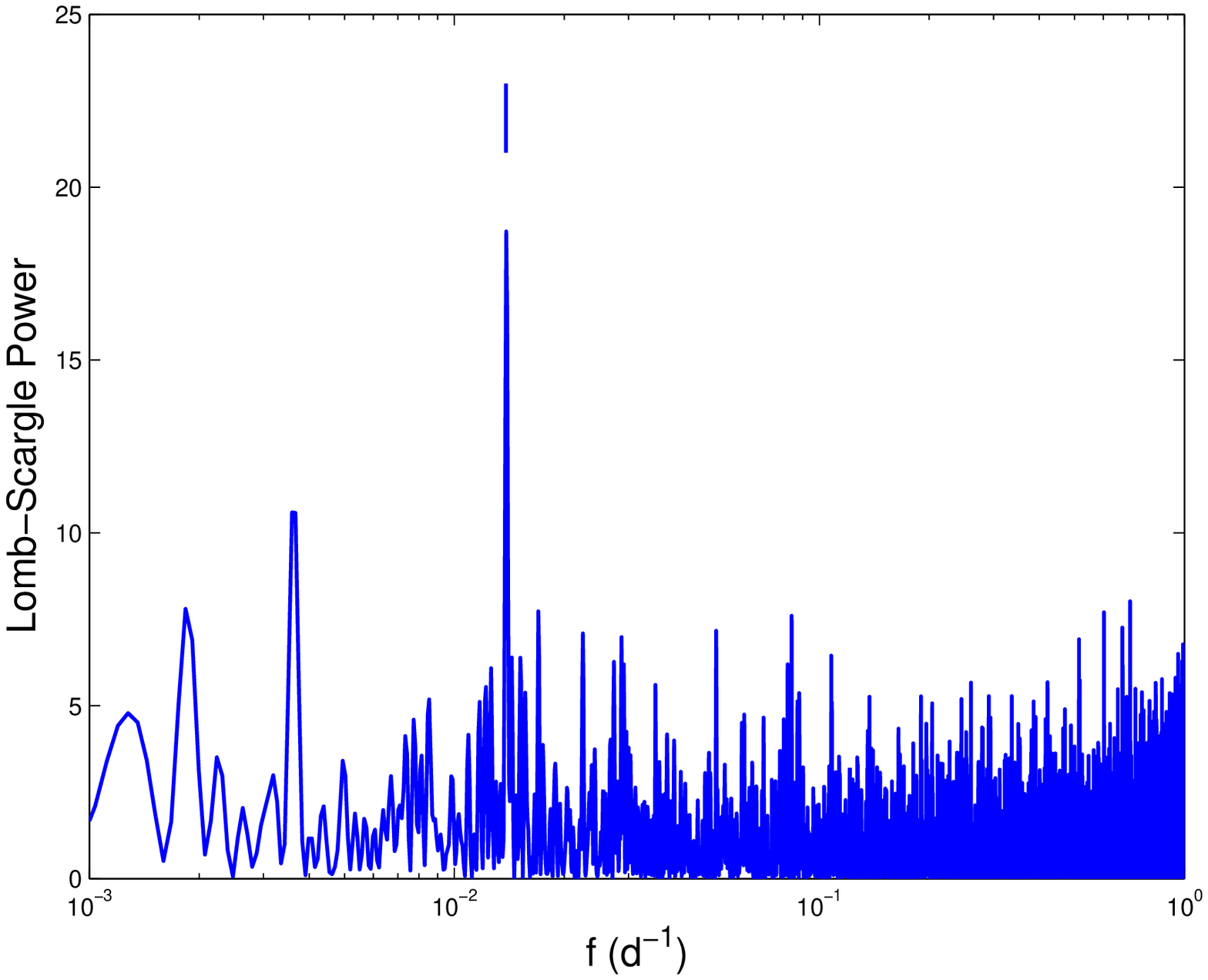}
\begin{figure}
\caption{The Lomb-Scargle periodogram for SAX~J1808.4$-$3658 in the
3--5 keV band made from 8.5 years of ASM data in 1-d time bins. The $\sim$ 72
d peak is indicated.}\label{lomb_saxj1808}
\end{figure}


\plotone{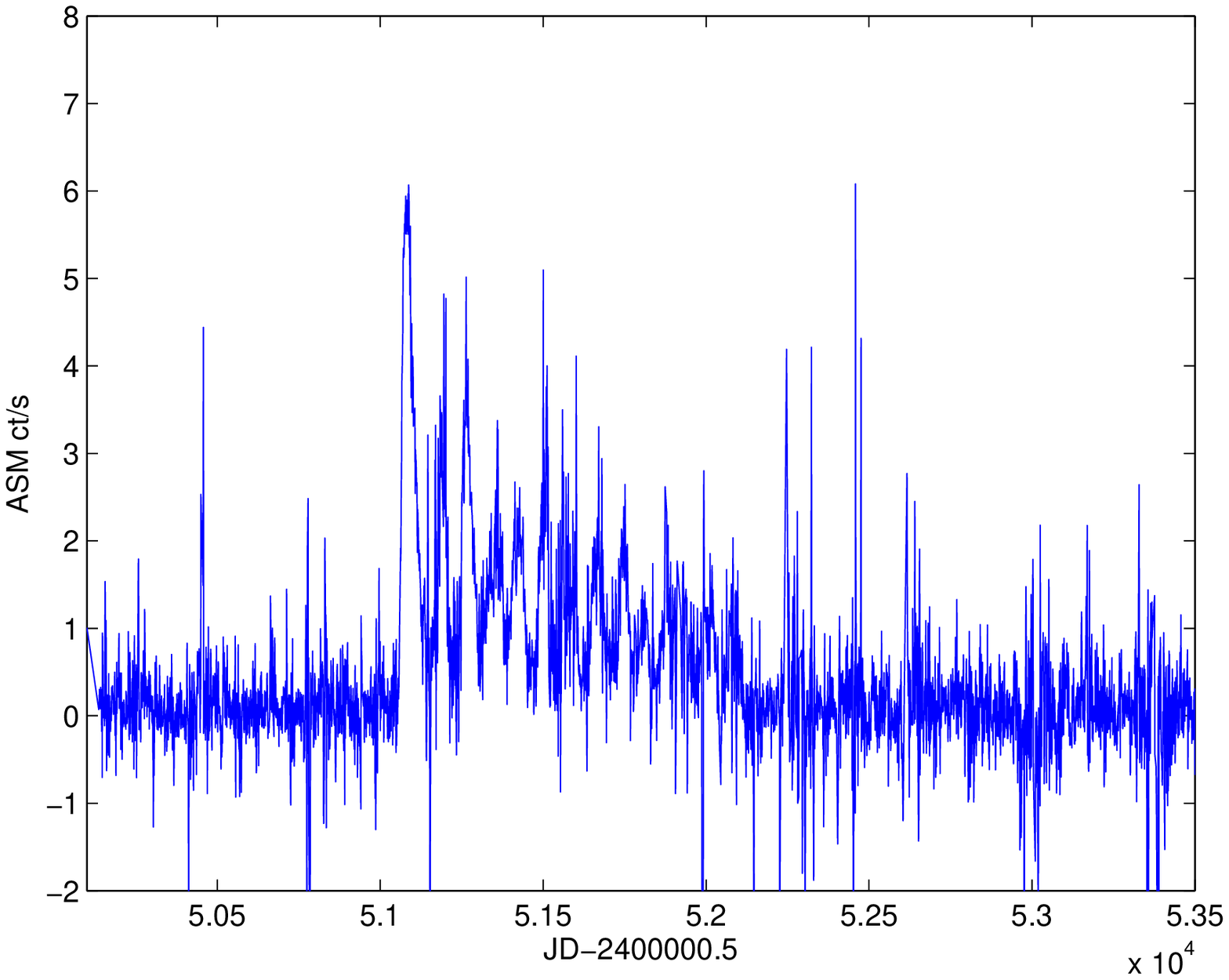}
\begin{figure}
\caption{The ASM light curve of X1942+274. Outbursts on
time scales of 80 d are apparent for the time interval of MJD 51000--52000.}\label{ltc_x1942}
\end{figure}

\plotone{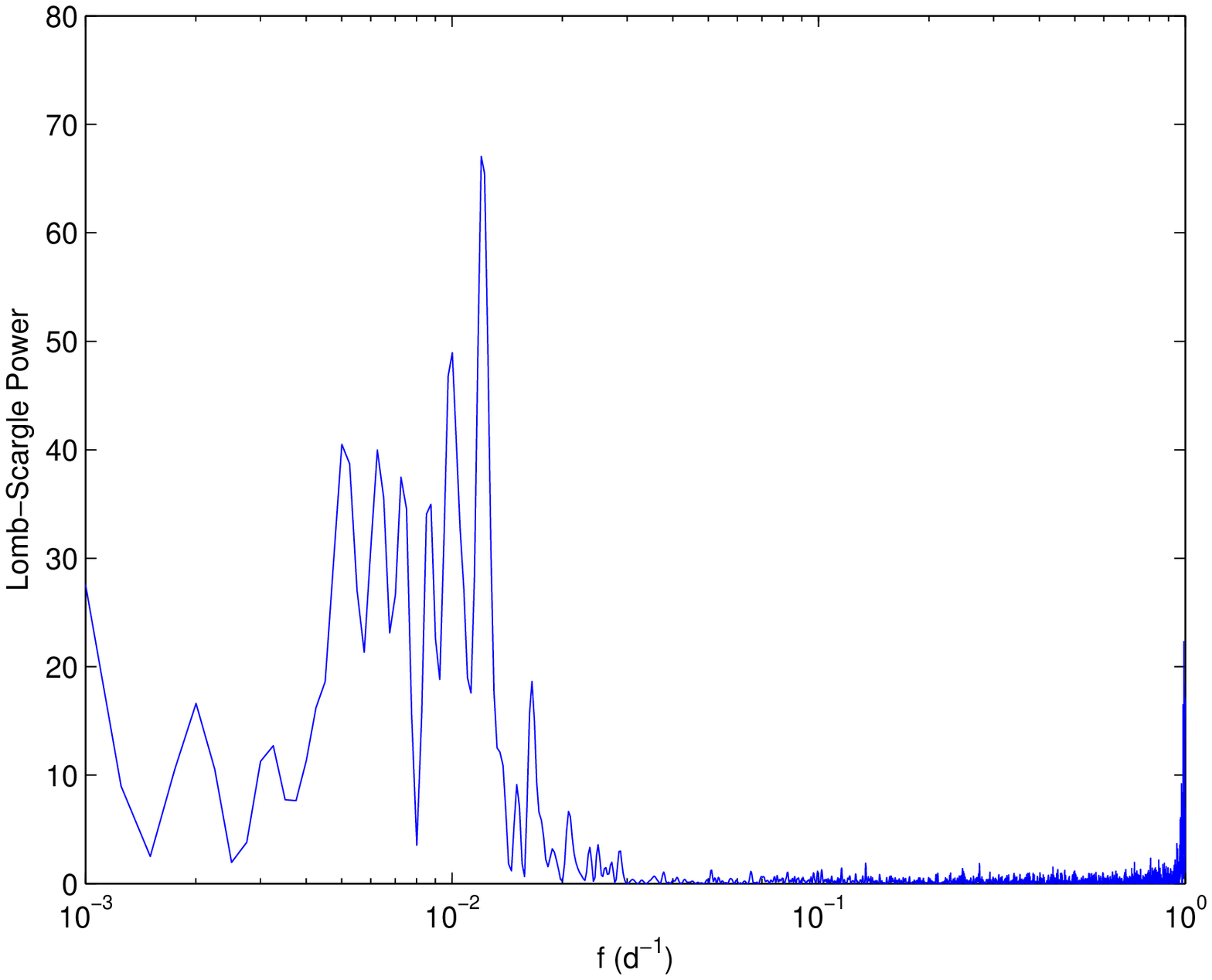}
\begin{figure}
\caption{The  Lomb-Scargle periodogram for X1942+274 in the 5--12 keV
  band for the time interval of MJD 51000-52000.}\label{power_x1942}
\end{figure}

\plotone{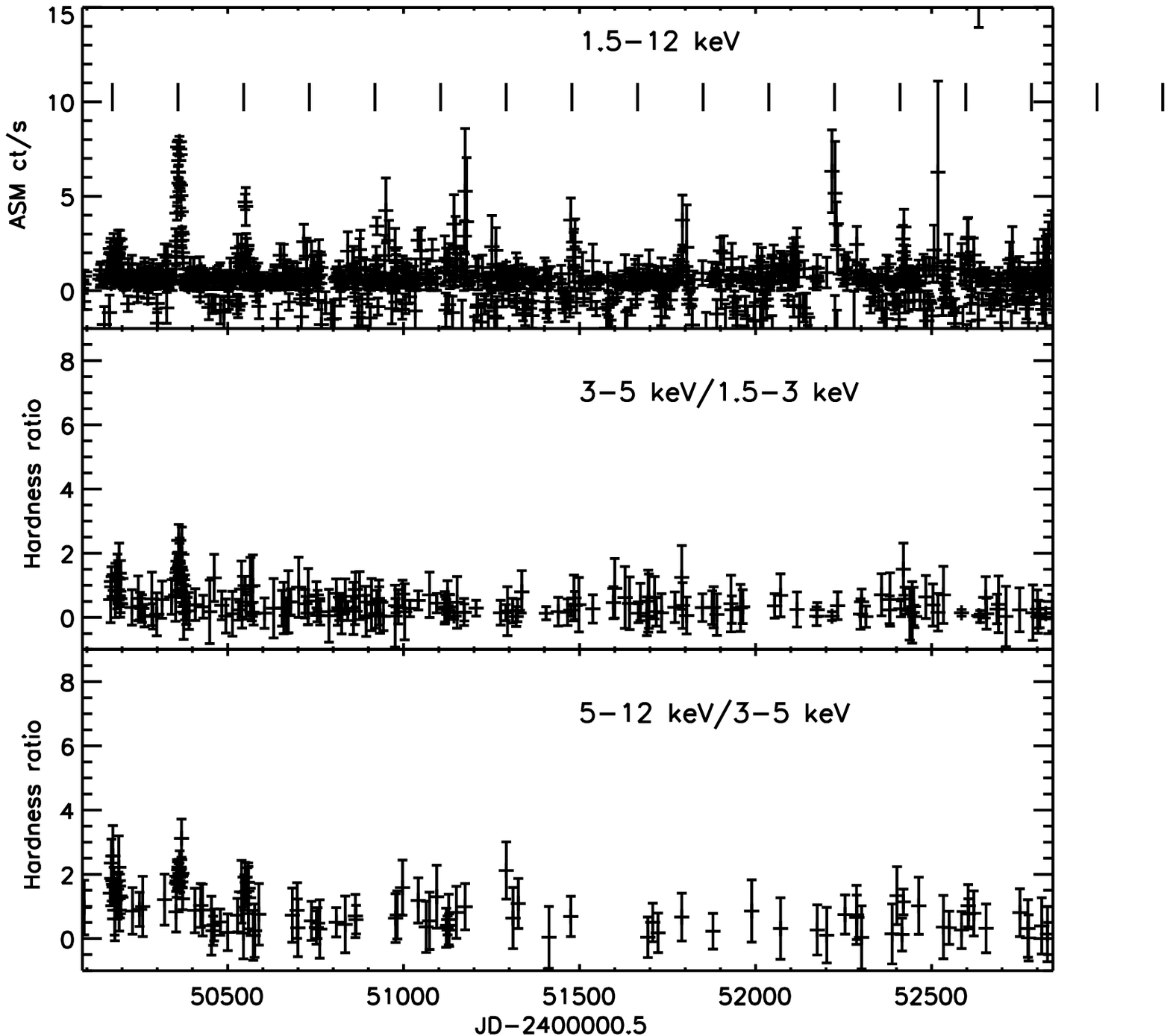}
\begin{figure}
\psfrag{Power}{Lomb-Scargle Periodogram}
\caption{The ASM light curve and hardness ratios of X1145$-$619. Intensity fluctuations on
time scales~$\sim$ 188 day intervals are labeled with short vertical lines.}\label{ltc_hr_x1145}
\end{figure}

\plotone{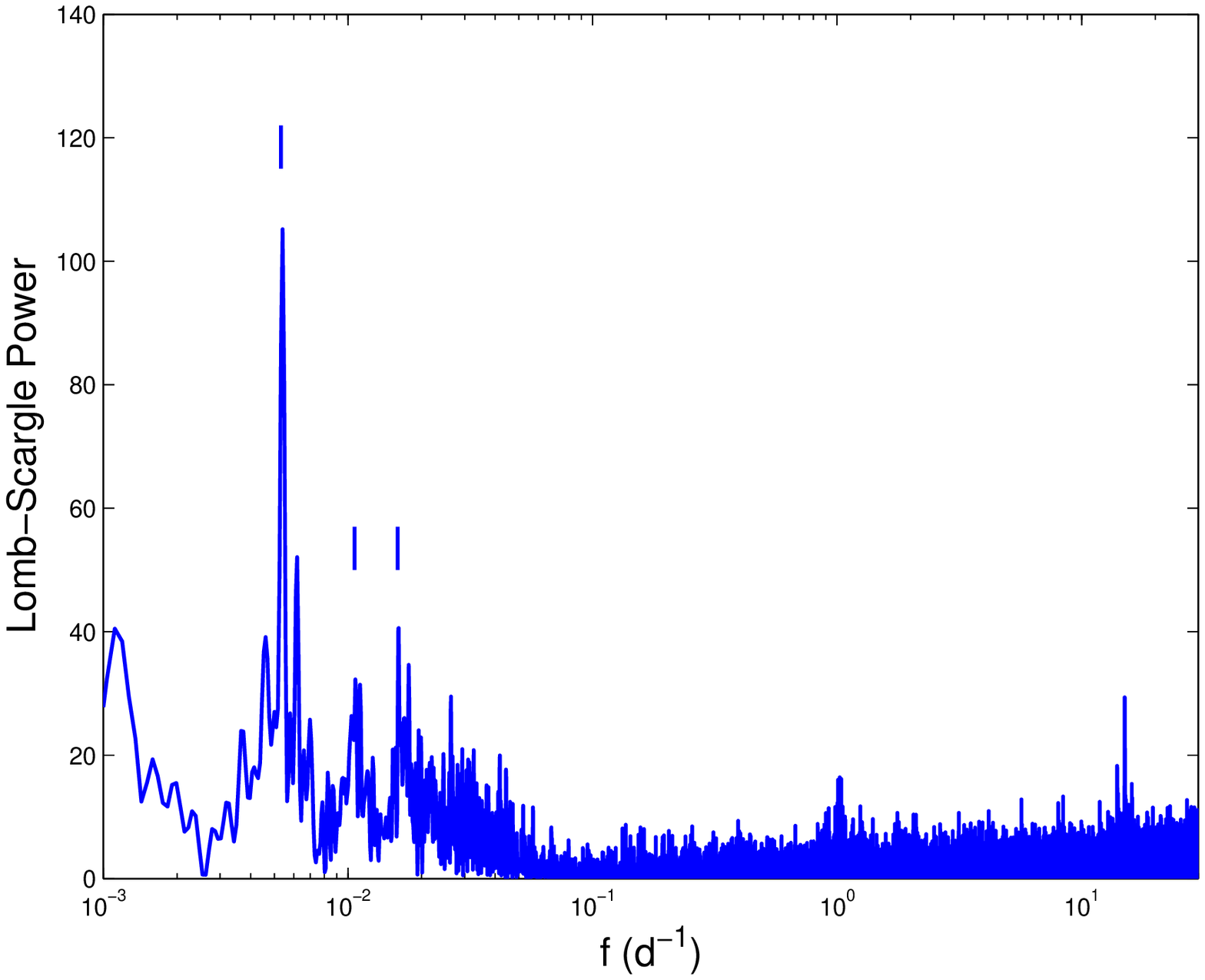}
\begin{figure}
\caption{Lomb-Scargle periodogram of X1145$-$619  in the 5--12 keV
  band with the 8.5 years of 
  ASM data in 90~s time bins.  The 
  $\sim$ 185 d period and its harmonic are indicated with short
  vertical lines. There are broad structures around the period and its
  harmonics.}\label{x1145_lomb_390}
\end{figure}

\pagebreak

\begin{table}
 \begin{center}
\footnotesize
\caption{Spin and Orbital Periodicities Detected with the  {\it RXTE}/ASM} 
\label{period_orb} 
\vspace*{0.4cm}
\begin{tabular}{ l l  l l l }
\hline
\hline
 Source name & Period  & FAP  & Energy Band &Comments \\
& & & (keV) & \\
\hline
\multicolumn{5}{l}{\footnotesize NS spin periods}\\
X0114+650  &  2.7421(2) h  &7e-5 $^a$              & 1.5--12  & HMXB/SG? \\ 
 X0352+30 (X~Per)   & 837.8(1) s & 3e-7  $^b$             & 1.5--12 & HMXB/Be \\
\hline
\multicolumn{5}{l}{\footnotesize Orbital periods (days): CVs  }\\
  AM Her            &  0.1289265(3)   & 1e-16 & 1.5--12 & Polar \\
\hline
\multicolumn{5}{l}{\footnotesize Orbital periods (in Units of days): LMXBs }\\
  EXO0748$-$676      & 0.1593375(6)   & 6e-7 & 3--5 & Eclipse \\
  X1658$-$298        & 0.296499(4) & 5e-4 $^c$ & 1.5--12  & Eclipse \\
  Her~X-1           & 1.70015(9)   & 5e-50  &5--12 & Eclipse \\
  X1822$-$371        &  0.2321091(6)     & 3e-57 &1.5--12& ADC  \\
  X2127+119 (M 15) & 0.713026(1) & 3e-8 $^c$ &3--5 & ADC  \\
  X1624$-$490        &0.86990(2)  & 1e-33& 5--12 & ADC, Dip\\
  X1916$-$053        & 0.03472969(5) &  1e-3 $^c$ & 1.5--12& Dip \\
  Cir~X-1            & 16.55(1)  & 5e-21 &1.5--3& Flares \&  dips \\
\hline
\multicolumn{5}{l}{\footnotesize  Orbital periods (days) :  HMXB supergiant systems} \\
  LMC~X-4            & 1.40840(6)  & 5e-8 & 5--12  & Eclipse \\
  Cen~X-3 & 2.08706(9)    & 5e-56 & 1.5--3 & Eclipse\\
  X1538$-$522 & 3.7284(3)  & 2e-81&1.5--12& Eclipse \\ 
  SMC~X-1 & 3.8921(4)   & 1e-50 & 5--12 & Eclipse \\
  Vela~X-1 &  8.965 (4) & 1e-115 &1.5--12 & Eclipse \\
  X1657$-$415 & 10.446(2)  & 3e-37 &5--12 & Eclipse \\ 
  Cyg~X-3 & 0.1996907(7) & 1e-195 & 5--12 & Wind ?\\ 
  X1700$-$377 &  3.4117(2)  & 1e-147 &3--5 & Wind  \\
  X1908+075 & 4.4005(4) & 7e-10 &1.5--12 & Wind? \\ 
  Cyg~X-1 & 5.6008(7) & 7e-12 $^d$   &1.5--12 & Wind  \\ 
  XTE~J1855$-$026   & 6.0752(8) & 6e-19 &5--12 & Wind ? \\ 
  X1907+097 &  8.375(2) &2e-26&5--12 &  Wind ?  \\ 
  X0114+650   & 11.599(5)   & 6e-12& 1.5--12  & Wind ? \\ 
  IGR J$19140+0951$ & 13.552(3) & 1e-18 & 5--12 & \\ 
  IGR J$00370+6122$ & 15.670(4) & 1e-14 & 1.5--12 & \\ 
  GX301-2 & 41.45(6) & $<$4e-211 &5--12 & Outburst/dip \\ 
\hline
\multicolumn{5}{l}{\footnotesize Orbital periods (days) : HMXB Be-star systems} \\
  X2206+543 & 9.561(3) &6e-12 &1.5--12& Outburst \\
  SAX~J2103.5+4545 & 12.673(4)  & 2e-6 & 1.5--12&  \\ 
  LSI+61303  & 26.2(1) &4e-3 $^e$ & 1.5--12 &  \\
  X0726$-$260   &  34.44(4)  &7e-14 &5--12 & Outburst \\ 
  EXO2030+375 & 46.14(6) & 1e-60 &5--12 & Outburst\\ 
  IGR J$11435-6109$ & 52.36(8) & 2e-2 & 5--12 & \\ 
  GRO~J2058+42   &  55.1(1) $^f$ & 2e-6 &1.5--12  & Outburst \\ 
  RX~J0812.4$-$3115    & 80.8(2) & 4e-5 $^d$  &5--12 &  Outburst \\ 
\hline
\end{tabular} \\[2pt]
\end{center}
\end{table}

\begin{table}
 \begin{center}
\small
\caption{Superorbital Periodicities Detected with the  {\it RXTE}/ASM} 
\label{period_long} 
\vspace*{0.4cm}
 \begin{tabular}{l l l l c l l l }
\hline
\hline
Source name & Period & FAP& Energy Band & Comments \\
 &  &  & (keV) &  \\
\hline
\multicolumn{5}{l}{\small Periodicities (days)} \\
LMC~X-4 & 30.28(2) & 2e-91 &5--12 &HMXB/P\\ 
X0114+650   & 30.75(3) & 5e-9 $^d$  & 5--12   & HMXB/SG? \\ 
Her~X-1 & 35.00(4) & 3e-20 &1.5--3&LMXB/P \\ 
XTE~J1716$-$389& 99.1(4) & 2e-6 $^d$ &1.5--12 &   LMXB(?)  \\ 
SS~433 &  162(1) & 3e-3 &5--12 &  HMXB/jet& \\ 
X1820$-$303  & 172(1)  & 6e-18 &5--12& LMXB  \\ 
\hline
\multicolumn{5}{l}{\small Quasi-periodicities (days)} \\
SMC~X-1 & 55.5(2) (50--70) & 3e-10& 1.5--12 & HMXB/P  \\
Cyg~X-2 & 68.9(2) (60-90) & 1e-2 $^d$ &1.5--12 & LMXB \\ 
GRS1747$-$312&  146.7(9) & 3e-6& 1.5--12& LMXB/NS\\
X1730$-$333& 217(2) &2e-5 $^d$  & 1.5--15& LMXB\\ 
LMC~X-3 & 453(17) (100-500) & 2e-3 $^d$ &5--12 &  HMXB\\ 
\hline
\end{tabular}
\end{center}
\end{table}

\clearpage
\newpage

\begin{table}
 \begin{center}
\small
\caption{Periodicities Detected with the  {\it RXTE}/ASM: Special Cases }
\label{period_special} 
\vspace*{0.4cm}
 \begin{tabular}{l l l l c l l l }
\hline
\hline
Source name & Period & FAP& Energy Band & Comments \\
 & (days) &  & (keV) &  \\
\hline
IRAS04575$-$7537 & 3.2324(3) & 5e-5 $^d$  & 1.5--12 & \\
X0115+634  & 24 & ($g$) &3--5 & HMXB/Be, Outburst  \\ 
SAX~J1808.4$-$3658 & 72.2(5) & 5e-5 $^d$ &3--5& LMXB/P \\ 
X1942+274     &78.7 & ($g$)  &5--12 &HMXB/Be, Outburst\\ 
	X1145$-$619 & 185 & ($g$) & 5--12 & Outburst\\
\hline
\end{tabular}
\end{center}

\vspace*{0.4cm} 
\addtocounter{footnote}{-3}
{$a$ } Detected in 0.4 years (MJD 50425.46-50552.84) \\ 
{$b$ } Detected with extended frequency range up to 120 cycles d$^{-1}$\\ 
{$c$ } FAP quoted from the first 6.5 years' data \\ 
{$d$ } FAP quoted from the first 8.6 years' 1-d average
data.\\ 
{$e$ } Detected  from the first 2.5 years' data \\  
{$f$ The ASM detected $\frac{1}{2}$  the orbital period.  } \\
{$g$ Highest peak in the original Lomb-Scargle periodogram, but not detected in the rescaled periodogram due to its broad peak. See discussions in subsection~\ref{special}  } 
\end{table} 

\end{document}